# Non-Termination Inference of Logic Programs

Etienne Payet and Fred Mesnard

IREMIA, Université de La Réunion, France

We present a static analysis technique for non-termination inference of logic programs. Our framework relies on an extension of the subsumption test, where some specific argument positions can be instantiated while others are generalized. We give syntactic criteria to statically identify such argument positions from the text of a program. Atomic left looping queries are generated bottom-up from selected subsets of the binary unfoldings of the program of interest. We propose a set of correct algorithms for automating the approach. Then, non-termination inference is tailored to attempt proofs of optimality of left termination conditions computed by a termination inference tool. An experimental evaluation is reported. When termination and non-termination analysis produce complementary results for a logic procedure, then with respect to the leftmost selection rule and the language used to describe sets of atomic queries, each analysis is optimal and together, they induce a *characterization* of the operational behavior of the logic procedure.

## 1 Introduction

Since the work of N. Lindenstrauss on TermiLog [20, 12], several automatic tools for termination checking (e.g. TALP [3]) or termination inference (e.g. cTI [25, 26] or TerminWeb [17]) are now available to the logic programmer. As the halting problem is undecidable for logic programs, such analyzers compute sufficient termination conditions implying left termination. In most works, only universal left termination is considered and termination conditions rely on a language for describing classes of atomic queries. The search tree associated to *any* (concrete) query satisfying a termination condition is guaranteed to be finite. When terms are abstracted using the *term-size* norm, the termination conditions are (disjunctions of) conjunctions of conditions of the form "the $i$-th argument is ground". Let us call this language $\mathcal{L}_{term}$.

In this report, which is based on an earlier conference paper [27], we present the first approach to non-termination inference tailored to attempt proofs of *optimality* of termination conditions at *verification time* for pure logic programs. The aim is to ensure the existence, for each class of atomic queries not covered by a termination condition, of *one* query from this class which leads to an infinite search tree when such a query is proved using any standard Prolog engine. We shall first present an analysis which computes classes of left looping queries, where any atomic query from such a class is guaranteed to lead to at least one infinite derivation under the usual left-to-right selection rule. Intuitively, we begin by computing looping queries from recursive binary clauses of the form $p(\ldots) \leftarrow p(\ldots)$. Then we try to add binary clauses of the form $q(\ldots) \leftarrow p(\ldots)$ to increase the set of looping queries. Finally by combining the result of non-termination inference with termination inference, for each predicate, we compute the set of modes for which the overall verification system has no information.

The main contributions of this work are:

- A new application of binary unfoldings to left loop inference. [16] introduced the binary

unfoldings of a logic program $P$ as a goal independent technique to transform $P$ into a possibly infinite set of binary clauses, which preserves the termination property [7] while abstracting the standard operational semantics. We present a correct algorithm to *construct* left looping classes of atomic goals, where such classes are computed bottom-up from selected subsets of the binary unfoldings of the analyzed program.

- A correct algorithm which, when combined with termination inference [23], may detect *optimal* left termination conditions expressed in $\mathcal{L}_{term}$ for logic programs. When termination and non-termination analysis produce complementary results for a logic procedure, then with respect to the leftmost selection rule and the language used to describe sets of atomic queries, each analysis is optimal and together, they induce a *characterization* of the operational behavior of the logic procedure.

- A report on the experimental evaluation we conduct. We have fully implemented termination and non-termination inference for logic programs. We have run the couple of analyzers on a set of classical logic programs, the sizes of which range from 2 to 177 clauses. The results of this experiment should help the reader to appreciate the value of the approach.

We organize the paper as follows: Section 2 presents the notations. In Section 3 we study loop inference for binary programs. We offer a full set of correct algorithms for non-termination inference in Section 4 and optimality proofs of termination conditions in Section 5. Finally, in Section 6, we discuss related works. The detailed proofs of the results can be found in Appendix B, at the end of the article.

## 2 Preliminaries

### 2.1 Functions

Let $E$ and $F$ be two sets. Then, $f : E \rightarrow F$ denotes that $f$ is a partial function from $E$ to $F$ and $f : E \rightarrowtail F$ denotes that $f$ is a function from $E$ to $F$. The *domain* of a partial function $f$ from $E$ to $F$ is denoted by $Dom(f)$ and is defined as: $Dom(f) = \{e \mid e \in E, \ f(e) \ \text{exists}\}$. Thus, if $f$ is a function from $E$ to $F$, then $Dom(f) = E$. Finally, if $f : E \rightarrow F$ is a partial function and $E'$ is a set, then $f|E'$ is the function from $Dom(f) \cap E'$ to $F$ such that for each $e \in Dom(f) \cap E'$, $f|E'$ maps $e$ to $f(e)$.

### 2.2 Logic Programming

We strictly adhere to the notations, definitions, and results presented in [1].

$N$ denotes the set of non-negative integers and for any $n \in N$, $[1, n]$ denotes the set $\{1, \dots, n\}$. If $n = 0$ then $[1, n] = \varnothing$.

From now on, we fix a language $\mathcal{L}$ of programs. We assume that $\mathcal{L}$ contains an infinite number of constant symbols. The set of relation symbols of $\mathcal{L}$ is $\Pi$, and we assume that each relation symbol $p$ has a *unique* arity, denoted $arity(p)$. $TU_{\mathcal{L}}$ (resp. $TB_{\mathcal{L}}$) denotes the set of all (ground and non ground) terms of $\mathcal{L}$ (resp. atoms of $\mathcal{L}$). A *query* is a finite sequence of atoms $A_1, \dots, A_n$ (where $n \geq 0$). When $n = 1$, we say that the query is *atomic*. Throughout this article, the variables of $\mathcal{L}$ are denoted by $X, Y, Z, \dots$, the constant symbols by $a, b, \dots$, the function symbols by $f, g, h, \dots$, the relation symbols by $p, q, r, \dots$, the atoms by $A, B, \dots$ and the queries by $Q, Q'$, $\dots$ or by $\mathbf{A}$, $\mathbf{B}$, $\dots$

Let $t$ be a term. Then $Var(t)$ denotes the set of variables occurring in $t$. This notation is extended to atoms, queries and clauses. Let $\theta := \{X_1/t_1, \dots, X_n/t_n\}$ be a substitution. We denote by $Dom(\theta)$ the set of variables $\{X_1, \dots, X_n\}$ and by $Ran(\theta)$ the set of variables appearing in $t_1, \dots, t_n$. We define $Var(\theta) = Dom(\theta) \cup Ran(\theta)$. Given a set of variables $V$, $\theta|V$ denotes the substitution obtained from $\theta$ by restricting its domain to $V$.

Let $t$ be a term and $\theta$ be a substitution. Then, the term $t\theta$ is called an *instance* of $t$. If $\theta$ is a *renaming* (*i.e.* a substitution that is a 1-1 and onto mapping from its domain to itself), then $t\theta$ is called a *variant* of $t$. Finally, $t$ is called *more general than* $t'$ if $t'$ is an instance of $t$.

A *logic program* is a finite set of definite clauses. In program examples, we use the ISO-Prolog syntax. Let $P$ be a logic program. Then $\Pi_P$ denotes the set of relation symbols appearing in $P$. In this paper, we only focus on left derivations *i.e.* we only consider the leftmost selection rule. Consider a non-empty query $B, \mathbf{C}$ and a clause $c$. Let $H \leftarrow \mathbf{B}$ be a variant of $c$ variable disjoint with $B, \mathbf{C}$ and assume that $B$ and $H$ unify. Let $\theta$ be an mgu of $B$ and $H$. Then $B, \mathbf{C} \stackrel{\theta}{\underset{c}{\Longrightarrow}} (\mathbf{B}, \mathbf{C})\theta$ is a *left derivation step* with $H \leftarrow \mathbf{B}$ as its *input clause*. If the substitution $\theta$ or the clause $c$ is irrelevant, we drop a reference to it.

Let $Q_0$ be a query. A maximal sequence $Q_0 \stackrel{\theta_1}{\underset{c_1}{\Longrightarrow}} Q_1 \stackrel{\theta_2}{\underset{c_2}{\Longrightarrow}} \cdots$ of left derivation steps is called a *left derivation* of $P \cup \{Q_0\}$ if $c_1, c_2, \ldots$ are clauses of $P$ and if the *standardization apart* condition holds, *i.e.* each input clause used is variable disjoint from the initial query $Q_0$ and from the mgu's and input clauses used at earlier steps. A finite left derivation may end up either with the empty query (then it is a *successful* left derivation) or with a non-empty query (then it is a *failed* left derivation). We say $Q_0$ *left loops* with respect to (w.r.t.) $P$ if there exists an infinite left derivation of $P \cup \{Q_0\}$. We write $Q \stackrel{+}{\underset{P}{\Longrightarrow}} Q'$ if there exists a finite non-empty prefix ending at $Q'$ of a left derivation of $P \cup \{Q\}$.

### 2.3 The Binary Unfoldings of a Logic Program

Let us present the main ideas about the *binary unfoldings* [16] of a logic program, borrowed from [7]. This technique transforms a logic program $P$ into a possibly infinite set of binary clauses. Intuitively, each generated *binary clause* $H \leftarrow B$ (where $B$ is either an atom or the atom *true* which denotes the empty query) specifies that, with respect to the original program $P$, a call to $H$ (or any of its instances) necessarily leads to a call to $B$ (or its corresponding instance).

More precisely, let $Q$ be an atomic query. Then $A$ is a *call* in a left derivation of $P \cup \{Q\}$ if $Q \stackrel{+}{\underset{P}{\Longrightarrow}} A, \mathbf{B}$. We denote by $calls_P(Q)$ the set of calls which occur in the left derivations of $P \cup \{Q\}$. The specialization of the goal independent semantics for call patterns for the left-to-right selection rule is given as the fixpoint of an operator $T_P^\beta$ over the domain of binary clauses, viewed modulo renaming. In the definition below, *id* denotes the set of all binary clauses of the form *true* $\leftarrow$ *true* or $p(X_1, \ldots, X_n) \leftarrow p(X_1, \ldots, X_n)$ for any $p \in \Pi_P$, where $arity(p) = n$.

$$T_P^\beta(X) = \left\{ (H \leftarrow B)\theta \,\middle|\, \begin{array}{l} c := H \leftarrow B_1, \ldots, B_m \in P, \ i \in [1, m], \\ \langle H_j \leftarrow true \rangle_{j=1}^{i-1} \in X \text{ renamed with fresh variables,} \\ H_i \leftarrow B \in X \cup id \text{ renamed with fresh variables,} \\ i < m \Rightarrow B \neq true \\ \theta = mgu(\langle B_1, \ldots, B_i \rangle, \langle H_1, \ldots, H_i \rangle) \end{array} \right\}$$

We define its powers as usual. It can be shown that the least fixpoint of this monotonic operator always exists and we set $bin\_unf(P) := lfp(T_P^\beta)$. Then the calls that occur in the left derivations of $P \cup \{Q\}$ can be characterized as follows: $calls_P(Q) = \{B\theta | H \leftarrow B \in bin\_unf(P), \theta = mgu(Q, H)\}$. This last property was one of the main initial motivations of the proposed abstract semantics, enabling logic programs optimizations. Similarly, $bin\_unf(P)$ gives a goal independent representation of the success patterns of $P$.

But we can extract more information from the binary unfoldings of a program $P$: universal left termination of an atomic query $Q$ with respect to $P$ is identical to universal termination of $Q$ with respect to $bin\_unf(P)$. Note that the selection rule is irrelevant for a binary program and an atomic query, as each subsequent query has at most one atom. The following result lies at the heart of Codish's approach to termination:

**Theorem 2.1** *[7] Let $P$ be a program and $Q$ an atomic query. Then $Q$ left loops with respect to $P$ iff $Q$ loops with respect to $bin\_unf(P)$.*

Notice that $bin\_unf(P)$ is a possibly infinite set of binary clauses. For this reason, in the algorithms of Section 4, we compute only the first *max* iterations of $T_P^\beta$ where *max* is a parameter of the analysis. As an immediate consequence of Theorem 2.1, assume that we detect that $Q$ loops with respect to a subset of the binary clauses of $T_P^\beta \uparrow i$, with $i \in N$. Then $Q$ loops with respect to $bin\_unf(P)$ hence $Q$ left loops with respect to $P$.

**Example 2.2** *Consider the following program $P$ (see [21], p. 56–58):*

```
p(X,Z) :- p(Y,Z),q(X,Y).     p(X,X).     q(a,b).
```

*The binary unfoldings of $P$ are:*

$$
\begin{aligned}
T_P^\beta \uparrow 0 &= \varnothing \\
T_P^\beta \uparrow 1 &= \{p(X,Z) \leftarrow p(Y,Z), p(X,X) \leftarrow true, q(a,b) \leftarrow true\} \cup T_P^\beta \uparrow 0 \\
T_P^\beta \uparrow 2 &= \{p(a,b) \leftarrow true, p(X,Y) \leftarrow q(X,Y)\} \cup T_P^\beta \uparrow 1 \\
T_P^\beta \uparrow 3 &= \{p(X,b) \leftarrow q(X,a), p(X,Z) \leftarrow q(Y,Z)\} \cup T_P^\beta \uparrow 2 \\
T_P^\beta \uparrow 4 &= \{p(X,b) \leftarrow q(Y,a)\} \cup T_P^\beta \uparrow 3 \\
T_P^\beta \uparrow 5 &= T_P^\beta \uparrow 4 = bin\_unf(P)
\end{aligned}
$$

*Let $Q := p(X,b)$. Note that $Q$ loops w.r.t. $T_P^\beta \uparrow 1$, hence it loops w.r.t. $bin\_unf(P)$. So $Q$ left loops w.r.t. $P$.* □

# 3 Loop Inference Using Filters

In this paper, we propose a mechanism that, given a logic program $P$, generates *at verification time* classes of atomic queries that left loop w.r.t. $P$. Our approach is completely based on the binary unfoldings of $P$ and relies on Theorem 2.1. It consists in computing a finite subset *BinProg* of $bin\_unf(P)$ and then in inferring a set of atomic queries that loop w.r.t. *BinProg*. By Theorem 2.1, these queries left loop w.r.t. $P$.

Hence, we reduce the problem of inferring looping atomic queries w.r.t. a logic program to that of inferring looping atomic queries w.r.t. a binary program. This is why in the sequel, our definitions, results and discussions mainly concentrate on binary programs only.

The central point of our method is the subsumption test, as the following lifting lemma, specialized for the leftmost selection rule, holds:

**Lemma 3.1** *(One Step Lifting, [1]) Let $Q \underset{c}{\Longrightarrow} Q_1$ be a left derivation step, $Q'$ be a query that is more general than $Q$ and $c'$ be a variant of $c$ variable disjoint with $Q'$. Then, there exists a query $Q_1'$ that is more general than $Q_1$ and such that $Q' \underset{c'}{\Longrightarrow} Q_1'$ with input clause $c'$.*

From this result, we derive:

**Corollary 3.2** *Let $c := H \leftarrow B$ be a binary clause. If $B$ is more general than $H$ then $H$ loops w.r.t. $\{c\}$.*

**Corollary 3.3** *Let $c := H \leftarrow B$ be a clause from a binary program BinProg. If $B$ loops w.r.t. BinProg then $H$ loops w.r.t. BinProg.*

These corollaries provide two sufficient conditions that can be used to design an incremental bottom-up mechanism that infers looping atomic queries. Given a binary program *BinProg*, it suffices to build the set $\mathcal{Q}$ of atomic queries consisting of the heads of the clauses whose body is more general than the head. By Corollary 3.2, the elements of $\mathcal{Q}$ loop w.r.t. *BinProg*. Then, by Corollary 3.3, the head of the clauses whose body is more general than an element of $\mathcal{Q}$ can safely been added to $\mathcal{Q}$ while retaining the property that every query in $\mathcal{Q}$ loops w.r.t. *BinProg*.

Notice that using this technique, we may not detect some looping queries. In [15], the authors show that there is no algorithm that, when given a right-linear binary recursive clause (*i.e.* a binary clause $p(\cdots) \leftarrow p(\cdots)$ such that all variables occur at most once in the body) and given an atomic query, always decides in a finite number of steps whether or not the resolution stops. In the case of a linear atomic query (*i.e.* an atomic query such that all variables occur at most once) however, the halting problem of derivations w.r.t. one binary clause is decidable [33, 13, 14].

It can be argued that the condition provided by Corollary 3.2 is rather weak because it fails at inferring looping queries in some simple cases. This is illustrated by the following example.

**Example 3.4** *Let $c$ be the clause $p(X) \leftarrow p(f(X))$. We have the infinite derivation:*

$$p(X) \underset{c}{\Longrightarrow} p(f(X)) \underset{c}{\Longrightarrow} p(f(f(X))) \underset{c}{\Longrightarrow} p(f(f(f(X)))) \cdots$$

*But, since the body of $c$ is not more general than its head, Corollary 3.2 does not allow to infer that $p(X)$ loops w.r.t. $\{c\}$.* □

In this section, we distinguish a special kind of argument positions that are "neutral" for derivation. Our goal is to extend the relation "is more general than" by, roughly, disregarding the predicate arguments whose position has been identified as neutral. Doing so, we aim at inferring more looping queries.

Intuitively, a set of predicate argument positions $\Delta$ is "Derivation Neutral" (DN for short) for a binary clause $c$ when the following holds. Let $Q$ be an atomic query and $Q'$ be a query obtained by replacing by any terms the predicate arguments in $Q$ whose position is in $\Delta$. If $Q \underset{c}{\Longrightarrow} Q_1$ then $Q' \underset{c}{\Longrightarrow} Q_1'$ where $Q_1'$ is more general than $Q_1$ up to the arguments whose position is in $\Delta$.

**Example 3.5** *(Example 3.4 continued) The predicate $p$ has only one argument position, so let us consider $\Delta = \langle p \mapsto \{1\} \rangle$ which distinguishes position 1 for predicate $p$. For any derivation step $p(s) \underset{c}{\Longrightarrow} p(s_1)$ if we replace $s$ by any term $t$ then there exists a derivation step $p(t) \underset{c}{\Longrightarrow} p(t_1)$. Notice that $p(t_1)$ is more general than $p(s_1)$ up to the argument of $p$. So, by the intuition described above, $\Delta$ is DN for $c$. Consequently, as in $c$ the body $p(f(X))$ is more general than the head $p(X)$ up to the argument of $p$ which is neutral, by an extended version of Corollary 3.2 there exists an infinite derivation of $\{c\} \cup \{p(X)\}$.* □

Let us give some more concrete examples of DN positions.

**Example 3.6** *The second argument position of the relation symbol append in the program* `APPEND`*:*

```
append([],Ys,Ys).                          % C1
append([X|Xs],Ys,[X|Zs]) :- append(Xs,Ys,Zs).   % C2
```

*is DN for* `C2`*. Notice that a very common programming technique called* accumulator passing *(see for instance e.g. [28], p. 21–25) always produces DN positions. A classical example of the accumulator passing technique is the following program* `REVERSE`*:*

```
reverse(L,R) :- rev(L,[],R).               % C1
rev([],R,R).                               % C2
rev([X|Xs],R0,R) :- rev(Xs,[X|R0],R).      % C3
```

*Concerning termination, we may ignore the second and the third argument of rev in the recursive clause* `C3` *while unfolding a query with this clause. Only the first argument can stop the unfolding.* □

But we can be even more precise. Instead of only identifying positions that can be totaly disregarded as in the above examples, we can try to identify positions where we can place any terms for which a given condition holds.

**Example 3.7** *Consider the clause $c := p(f(X)) \leftarrow p(f(f(X)))$. If we mean by a DN position a position where we can place any terms, then the argument position of $p$ is not DN for $c$. This is because, for example, we have the derivation step $p(X) \underset{c}{\Longrightarrow} p(f(f(X_1)))$ but if we replace $X$ by $g(X)$ then there is no derivation step of $\{c\} \cup \{p(g(X))\}$. However, if we mean by a DN position a position where we can place any instances of $f(X)$, then the argument position of $p$ is DN for $c$.* □

In the sequel of the section, we define more precisely DN positions as positions where we can place any terms satisfying certain conditions identified by "filters". We use filters to present an extension of the relation "is more general than" and we propose an extended version of Corollary 3.2. We offer two syntactic conditions of increasing power for easily identifying DN positions from mere inspection of the text of a logic program. The practical impact of such filters will be tackled in Section 5.

## 3.1 Filters

Let us first introduce the notion of a *filter*. We use filters in order to distinguish atoms, some arguments of which satisfy a given condition. A condition upon atom arguments, *i.e.* terms, can be defined as a function in the following way.

**Definition 3.8** *(Term-condition) A* term-condition *is a function from the set of terms $TU_{\mathcal{L}}$ to $\{\mathtt{true}, \mathtt{false}\}$.*

**Example 3.9** *The following functions are term-conditions.*

$$
\begin{aligned}
f_{true} : \quad TU_{\mathcal{L}} \quad &\rightarrowtail \quad \{\mathtt{true}, \mathtt{false}\} \\
t \quad &\mapsto \quad \mathtt{true}
\end{aligned}
$$

$$
\begin{aligned}
f_1 : \quad TU_{\mathcal{L}} \quad &\rightarrowtail \quad \{\mathtt{true}, \mathtt{false}\} \\
t \quad &\mapsto \quad \mathtt{true} \text{ iff } t \text{ is an instance of } [X|Y]
\end{aligned}
$$

$$
\begin{aligned}
f_2 : \quad TU_{\mathcal{L}} \quad &\rightarrowtail \quad \{\mathtt{true}, \mathtt{false}\} \\
t \quad &\mapsto \quad \mathtt{true} \text{ iff } t \text{ unifies with } h(a, X)
\end{aligned}
$$

□

Notice that a term-condition might give distinct results for two terms which are equal modulo renaming. For instance $f_2(X) = \mathtt{false}$ and $f_2(Y) = \mathtt{true}$. However, in Definition 3.12 below, we will only consider *variant independent* term-conditions.

**Definition 3.10** *(Variant Independent Term-Condition) A term-condition $f$ is* variant independent *if, for every term $t$, $f(t) = \mathtt{true}$ implies that $f(t') = \mathtt{true}$ for every variant $t'$ of $t$.*

**Example 3.11** *(Example 3.9 continued) $f_{true}$ and $f_1$ are variant independent while $f_2$ is not.* □

We restrict the class of term-conditions to that of variant independent ones because we want to extend the relation "is more general than" so that if an atom $A$ is linked to an atom $B$ by the extended relation, then every variant of $A$ is also linked to $B$ (see Proposition 3.16 below). This will be essential to establish the forthcoming main Proposition 3.20 which is an extension of Corollary 3.2. Now we can define what we exactly mean by a filter.

**Definition 3.12** *(Filter) A* filter*, denoted by $\Delta$, is a function from $\Pi$ such that: for each $p \in \Pi$, $\Delta(p)$ is a partial function from $[1, arity(p)]$ to the set of variant independent term-conditions.*

**Example 3.13** *(Example 3.9 continued) Let $p$ be a relation symbol whose arity equals 3. The filter $\Delta$ which maps $p$ to the function $\langle 1 \mapsto f_{true}, \ 2 \mapsto f_1 \rangle$ and any $q \in \Pi \setminus \{p\}$ to $\langle \rangle$ is noted $\Delta := \langle \ p \mapsto \langle 1 \mapsto f_{true}, \ 2 \mapsto f_1 \rangle \ \rangle$.* □

## 3.2 Extension of the Relation "Is More General Than"

Given a filter $\Delta$, the relation "is more general than" can be extended in the following way: an atom $A := p(\cdots)$ is $\Delta$-more general than $B := p(\cdots)$ if the "is more general than" requirement holds for those arguments of $A$ whose position is not in the domain of $\Delta(p)$ while the other arguments satisfy their associated term-condition.

**Definition 3.14** *($\Delta$-more general) Let $\Delta$ be a filter and $A$ and $B$ be two atoms.*

- *Let $\eta$ be a substitution. Then $A$ is $\Delta$-more general than $B$ for $\eta$ if:*

$$
\begin{cases}
A = p(s_1, \ldots, s_n) \\
B = p(t_1, \ldots, t_n) \\
\forall i \in [1, n] \setminus Dom(\Delta(p)), \ t_i = s_i \eta \\
\forall i \in Dom(\Delta(p)), \ \Delta(p)(i)(s_i) = \mathtt{true}.
\end{cases}
$$

- *$A$ is $\Delta$-more general than $B$ if there exists a substitution $\eta$ s.t. $A$ is $\Delta$-more general than $B$ for $\eta$.*

*An atomic query $Q$ is $\Delta$-more general than an atomic query $Q'$ if either $Q$ and $Q'$ are both empty or $Q$ contains the atom $A$, $Q'$ contains the atom $B$ and $A$ is $\Delta$-more general than $B$.*

**Example 3.15** *(Example 3.13 continued) Let*

$$
\begin{aligned}
A &:= p( \quad b \ , \quad X \quad , \ h(a, X) \quad ) \\
B &:= p( \quad a \ , \ [a|b] \ , \quad X \quad ) \\
C &:= p( \quad a \ , \ [a|b] \ , \ h(Y, b) \quad ) .
\end{aligned}
$$

*Then, $A$ is not $\Delta$-more general than $B$ and $C$ because, for instance, its second argument $X$ is not an instance of $[X|Y]$ as required by $f_1$. On the other hand, $B$ is $\Delta$-more general than $A$ for the substitution $\{X/h(a, X)\}$ and $B$ is $\Delta$-more general than $C$ for the substitution $\{X/h(Y, b)\}$. Finally, $C$ is not $\Delta$-more general than $A$ because $h(Y, b)$ is not more general than $h(a, X)$ and $C$ is not $\Delta$-more general than $B$ because $h(Y, b)$ is not more general than $X$.* □

As in a filter the term-conditions are variant independent, we get the following proposition.

**Proposition 3.16** *Let $\Delta$ be a filter and $A$ and $B$ be two atoms. If $A$ is $\Delta$-more general than $B$ then every variant of $A$ is $\Delta$-more general than $B$.*

The next proposition states an intuitive result:

**Proposition 3.17** *Let $\Delta$ be a filter and $A$ and $B$ be two atoms. Then $A$ is $\Delta$-more general than $B$ if and only if there exists a substitution $\eta$ such that $Var(\eta) \subseteq Var(A, B)$ and $A$ is $\Delta$-more general than $B$ for $\eta$.*

## 3.3 Derivation Neutral Filters: Operational Definition

In the sequel of this paper, we focus on "derivation neutral" filters. The name "derivation neutral" stems from the fact that in any derivation of an atomic query $Q$, the arguments of $Q$ whose position is distinguished by such a filter can be safely replaced by any terms satisfying the associated term-condition. Such a replacement does not modify the derivation process.

**Definition 3.18** *(Derivation Neutral) Let $\Delta$ be a filter and $c$ be a binary clause. We say that $\Delta$ is DN for $c$ if for each derivation step $Q \underset{c}{\Longrightarrow} Q_1$ where $Q$ is an atomic query, for each $Q'$ that is $\Delta$-more general than $Q$ and for each variant $c'$ of $c$ variable disjoint with $Q'$, there exists a query $Q_1'$ that is $\Delta$-more general than $Q_1$ and such that $Q' \underset{c}{\Longrightarrow} Q_1'$ with input clause $c'$. This definition is extended to binary programs: $\Delta$ is DN for $P$ if it is DN for each clause of $P$.*

**Example 3.19** *The following examples illustrate the previous definition.*

- *Let us reconsider the program* APPEND *from Example 3.6 with the term-condition* $f_{true}$ *defined in Example 3.9 and the filter* $\Delta := \langle append \mapsto \langle 2 \mapsto f_{true} \rangle \rangle$. $\Delta$ *is DN for* C2. *However,* $\Delta$ *is not DN for* APPEND *because it is not DN for* C1.

- *Consider the following clause:*

  merge([X|Xs],[Y|Ys],[X|Zs]) :- merge(Xs,[Y|Ys],Zs).

  *The filter* $\langle merge \mapsto \langle 2 \mapsto f_1 \rangle \rangle$, *where the term-condition* $f_1$ *is defined in Example 3.9, is DN for this clause.*

*In the next subsection, we present some syntactic criteria for identifying correct DN filters. For proving that the above filters are indeed DN, we will just check that they actually fulfill these syntactic criteria that are sufficient conditions.* □

Derivation neutral filters lead to the following extended version of Corollary 3.2 (take $\Delta$ such that for any $p$, $\Delta(p)$ is a function whose domain is empty):

**Proposition 3.20** *Let* $c := H \leftarrow B$ *be a binary clause and* $\Delta$ *be a filter that is DN for* $c$. *If* $B$ *is* $\Delta$-more general than $H$ then $H$ loops w.r.t. $\{c\}$.

We point out that the above results remain valid when the program under consideration is restricted to its set of clauses used in the derivation steps. For instance, although the filter $\Delta$ of Example 3.19 is not DN for APPEND, it will help us to construct queries which loop w.r.t. C2. Such queries also loop w.r.t. APPEND.

Notice that lifting lemmas are used in the literature to prove completeness of SLD-resolution. As Definition 3.18 corresponds to an extended version of the One Step Lifting Lemma 3.1, it may be worth to investigate its consequences from the model theoretic point of view.

First of all, a filter may be used to "expand" atoms by replacing every argument whose position is distinguished by any term that satisfies the associated term-condition.

**Definition 3.21** *Let* $\Delta$ *be a filter and* $A$ *be an atom. The expansion of* $A$ *w.r.t.* $\Delta$, *denoted* $A_{\uparrow\Delta}$, *is the set defined as*

$$A_{\uparrow\Delta} \stackrel{def}{=} \{A\} \cup \{B \in TB_{\mathcal{L}} \mid B \text{ is } \Delta\text{-more general than } A \text{ for } \epsilon\}$$

*where* $\epsilon$ *denotes the empty substitution.*

Notice that in this definition, we do not necessary have the inclusion

$$\{A\} \subseteq \{B \in TB_{\mathcal{L}} \mid B \text{ is } \Delta\text{-more general than A for } \epsilon\} .$$

For instance, suppose that $A := p(f(X))$ and that $\Delta$ maps $p$ to the function $\langle 1 \mapsto f \rangle$ where $f$ is the term-condition mapping any term $t$ to true iff $t$ is an instance of $g(X)$. Then

$$\{B \in TB_{\mathcal{L}} \mid B \text{ is } \Delta\text{-more general than A}\} = \{p(t) \mid t \text{ is an instance of } g(X)\}$$

with $A \notin \{p(t) \mid t \text{ is an instance of } g(X)\}$.

Term interpretations in the context of logic programming were first introduced in [6] and further investigated in [11] and then in [22]. A term interpretation for $\mathcal{L}$ is identified with a (possibly empty) subset of the term base $TB_{\mathcal{L}}$. So, as for atoms, a term interpretation can be expanded by a filter.

**Definition 3.22** *Let* $\Delta$ *be a filter and* $I$ *be a term interpretation for* $\mathcal{L}$. *Then* $I_{\uparrow\Delta}$ *is the term interpretation for* $\mathcal{L}$ *defined as:*

$$I_{\uparrow\Delta} \stackrel{def}{=} \bigcup_{A \in I} A_{\uparrow\Delta} .$$

For any logic program $P$, we denote by $\mathcal{C}(P)$ its least term model.

**Theorem 3.23** *Let $P$ be a binary program and $\Delta$ be a DN filter for $P$. Then $\mathcal{C}(P)_{\uparrow\Delta} = \mathcal{C}(P)$.*

*Proof.* The inclusion $\mathcal{C}(P) \subseteq \mathcal{C}(P)_{\uparrow\Delta}$ is straightforward so let us concentrate on the other one *i.e.* $\mathcal{C}(P)_{\uparrow\Delta} \subseteq \mathcal{C}(P)$. Let $A' \in \mathcal{C}(P)_{\uparrow\Delta}$. Then there exists $A \in \mathcal{C}(P)$ such that $A' \in A_{\uparrow\Delta}$. A well known result states:

$$\mathcal{C}(P) = \{B \in TB_{\mathcal{L}} \mid \text{there exists a successful derivation of } P \cup \{B\}\} \tag{1}$$

Consequently, there exists a successful derivation $\xi$ of $P \cup \{A\}$. Therefore, by successively applying Definition 3.18 to each step of $\xi$, one construct a successful derivation of $A'$. So by (1) $A' \in \mathcal{C}(P)$. $\quad\square$

## 3.4 Some Particular DN Filters

In this section, we provide two sufficient syntactic conditions for identifying DN filters.

### 3.4.1 DN Sets of Positions

The first instance we consider corresponds to filters, the associated term-conditions of which are all equal to $f_{true}$ (see Example 3.9). Within such a context, as the term-conditions are fixed, each filter $\Delta$ is uniquely determined by the domains of the partial functions $\Delta(p)$ for $p \in \Pi$. Hence the following definition.

**Definition 3.24** *(Set of Positions) A set of positions, denoted by $\tau$, is a function from $\Pi$ to $2^N$ such that: for each $p \in \Pi$, $\tau(p)$ is a subset of $[1, arity(p)]$.*

**Example 3.25** *Let append and append3 be two relation symbols. Assume that $arity(append) = 3$ and $arity(append3) = 4$. Then $\tau := \langle\ append \mapsto \{2\},\ append3 \mapsto \{2, 3, 4\}\ \rangle$ is a set of positions.* $\square$

Not surprisingly, the filter that is generated by a set of positions is defined as follows.

**Definition 3.26** *(Associated Filter) Let $\tau$ be a set of positions and $f_{true}$ be the term-condition defined in Example 3.9. The filter $\Delta[\tau]$ defined as:*

$$\text{for each } p \in \Pi,\ \Delta[\tau](p) \text{ is the function from } \tau(p) \text{ to } \{f_{true}\}$$

*is called the filter associated to $\tau$.*

**Example 3.27** *(Example 3.25 continued) The filter associated to $\tau$ is*

$$\Delta[\tau] := \quad \langle append \mapsto \langle 2 \mapsto f_{true}\rangle, append3 \mapsto \langle 2 \mapsto f_{true}, 3 \mapsto f_{true}, 4 \mapsto f_{true}\rangle\rangle.$$

$\square$

Now we define a particular kind of sets of positions. These are named after "DN" because, as stated by Theorem 3.30 below, they generate DN filters.

**Definition 3.28** *(DN Set of Positions) Let $\tau$ be a set of positions. We say that $\tau$ is DN for a binary clause $p(s_1, \ldots, s_n) \leftarrow q(t_1, \ldots, t_m)$ if:*

$$\forall i \in \tau(p), \left\{\begin{array}{l} s_i \text{ is a variable} \\ s_i \text{ occurs only once in } p(s_1, \ldots, s_n) \\ \forall j \in [1, m],\ s_i \in Var(t_j) \Rightarrow j \in \tau(q)\ . \end{array}\right.$$

*A set of positions is* DN *for a binary program $P$ if it is DN for each clause of $P$.*

The intuition of Definition 3.28 is the following. If for instance we have a clause $c :=$ $p(X, Y, f(Z)) \leftarrow p(g(Y, Z), X, Z)$ then in the first two positions of $p$ we can put any terms and get a derivation step w.r.t. $c$ because the first two arguments of the head of $c$ are variables that appear exactly once in the head. Moreover, $X$ and $Y$ of the head reappear in the body but again only in the first two positions of $p$. So, if we have a derivation step $p(s_1, s_2, s_3) \underset{c}{\Longrightarrow} p(t_1, t_2, t_3)$, we can replace $s_1$ and $s_2$ by any terms $s_1'$ and $s_2'$ and get another derivation step $p(s_1', s_2', s_3) \underset{c}{\Longrightarrow} p(t_1', t_2', t_3')$ where $t_3'$ is the same as $t_3$ up to variable names.

**Example 3.29** *(Example 3.25 continued) $\tau$ is DN for the program:*

```
append([X|Xs],Ys,[X|Zs]) :- append(Xs,Ys,Zs).
append3(Xs,Ys,Zs,Ts) :- append(Xs,Ys,Us).
```

*which is a subset of the binary unfoldings of the program* `APPEND3`*:*

```
append([],Ys,Ys).
append([X|Xs],Ys,[X|Zs]) :- append(Xs,Ys,Zs).
append3(Xs,Ys,Zs,Ts) :- append(Xs,Ys,Us), append(Us,Zs,Ts).
```

$\square$

DN sets of positions generate DN filters.

**Theorem 3.30** *Let $\tau$ be a DN set of positions for a binary program $P$. Then $\Delta[\tau]$ is DN for $P$.*

*Proof.* As we will see in Section 3.4.2, this theorem is a particular case of Theorem 3.39.

Notice that the set of DN sets of positions of any binary program $P$ is not empty because, by Definition 3.28, $\tau_0 := \langle p \mapsto \varnothing \mid p \in \Pi \rangle$ is DN for $P$. Moreover, an atom $A$ is $\Delta[\tau_0]$-more general than an atom $B$ iff $A$ is more general than $B$.

### 3.4.2 DN Sets of Positions with Associated Terms

Now we consider another instance of Definition 3.18. As we will see, it is more general than the previous one. It corresponds to filters whose associated term-conditions have all the form "is an instance of $t$" where $t$ is a term that uniquely determines the term-condition. Notice that such term-conditions are variant independent, so it makes sense to consider such filters. Hence the following definition.

**Definition 3.31** *(Sets of Positions with Associated Terms) A set of positions with associated terms, denoted by $\tau^+$, is a function from $\Pi$ such that: for each $p \in \Pi$, $\tau^+(p)$ is a partial function from $[1, arity(p)]$ to $TU_{\mathcal{L}}$.*

**Example 3.32** *Let $p$ and $q$ be two relation symbols whose arity is 2. Then*

$$\tau^+ := \ \langle p \mapsto \langle 2 \mapsto X \rangle, \ q \mapsto \langle 2 \mapsto g(X) \rangle \ \rangle$$

*is a set of positions with associated terms.* $\square$

The filter that is generated by a set of positions with associated terms is defined as follows.

**Definition 3.33** *(Associated Filter) Let $\tau^+$ be a set of positions with associated terms. The filter associated to $\tau^+$, denoted by $\Delta[\tau^+]$, is defined as: for each $p \in \Pi$, $\Delta[\tau^+](p)$ is the function*

$$
\begin{aligned}
Dom(\tau^+(p)) \quad &\rightarrowtail \quad \textit{The set of term-conditions} \\
i \quad &\mapsto \quad \left\{ \begin{array}{l} TU_{\mathcal{L}} \ \rightarrowtail \ \{\texttt{true}, \texttt{false}\} \\ \quad t \quad \mapsto \quad \texttt{true} \textit{ iff } t \textit{ is an instance of } \tau^+(p)(i) \end{array} \right.
\end{aligned}
$$

**Example 3.34** *(Example 3.32 continued) The filter associated to $\tau^+$ is*

$$\Delta[\tau^+] := \ \langle\, p \mapsto \langle 2 \mapsto f_1\rangle,\ q \mapsto \langle 2 \mapsto f_2\rangle \,\rangle$$

*where*

$$
\begin{aligned}
f_1 : \quad & TU_\mathcal{L} \ \rightarrowtail \ \{\texttt{true}, \texttt{false}\} \\
& t \ \mapsto \ \texttt{true} \textit{ iff } t \textit{ is an instance of } X
\end{aligned}
$$

$$
\begin{aligned}
f_2 : \quad & TU_\mathcal{L} \ \rightarrowtail \ \{\texttt{true}, \texttt{false}\} \\
& t \ \mapsto \ \texttt{true} \textit{ iff } t \textit{ is an instance of } g(X)
\end{aligned}
$$

□

As for sets of positions, we define a special kind of sets of positions with associated terms.

**Definition 3.35** *(DN Sets of Positions with Associated Terms) Let $\tau^+$ be a set of positions with associated terms. We say that $\tau^+$ is DN for a binary clause $p(s_1, \ldots, s_n) \leftarrow q(t_1, \ldots, t_m)$ if these conditions hold:*

- **(DN1)** $\forall i \in Dom(\tau^+(p))$, $\forall j \in [1, n] \setminus \{i\}$: $Var(s_i) \cap Var(s_j) = \varnothing$,

- **(DN2)** $\forall \langle i \mapsto u_i\rangle \in \tau^+(p)$: $s_i$ *is more general than* $u_i$,

- **(DN3)** $\forall \langle j \mapsto u_j\rangle \in \tau^+(q)$: $t_j$ *is an instance of* $u_j$,

- **(DN4)** $\forall i \in Dom(\tau^+(p))$, $\forall j \notin Dom(\tau^+(q))$: $Var(s_i) \cap Var(t_j) = \varnothing$.

*A set of positions with associated terms is DN for a binary program $P$ if it is DN for each clause of $P$.*

This definition says that any $s_i$ where $i$ is in the domain of $\tau^+(p)$ (*i.e.* position $i$ is distinguished by $\tau^+$): **(DN1)** does not share its variables with the other arguments of the head, **(DN2)** is more general than the term $u_i$ that $i$ is mapped to by $\tau^+(p)$, **(DN4)** distributes its variables to some $t_j$ such that $j$ is in the domain of $\tau^+(q)$ (*i.e.* position $j$ is distinguished by $\tau^+$). Moreover, **(DN3)** says that any $t_j$, where $j$ is distinguished by $\tau^+$, is such that $t_j$ is an instance of the term $u_j$ that $j$ is mapped to by $\tau^+(q)$.

**Example 3.36** *(Example 3.32 continued) $\tau^+$ is DN for the following program:*

```
p(f(X),Y) :- q(X,g(X)).
q(a,g(X)) :- q(a,g(b)).
```

The preceding notion is closed under renaming:

**Proposition 3.37** *Let $c$ be a binary clause and $\tau^+$ be a set of positions with associated terms that is DN for $c$. Then $\tau^+$ is DN for every variant of $c$.*

Notice that a set of positions is a particular set of positions with associated terms in the following sense.

**Proposition 3.38** *Let $\tau$ be a set of positions and $X$ be a variable. Let $\tau^+$ be the set of positions with associated terms defined as: for each $p \in \Pi$, $\tau^+(p) := (\, \tau(p) \rightarrowtail \{X\} \,)$. Then, the following holds.*

1. *An atom $A$ is $\Delta[\tau]$-more general than an atom $B$ iff $A$ is $\Delta[\tau^+]$-more general than $B$.*

2. *For any binary clause $c$, $\tau$ is DN for $c$ iff $\tau^+$ is DN for $c$.*

*Proof.* A proof follows from these remarks.

- Item 1 is a direct consequence of the definition of "Δ-more general" (see Definition 3.14) and the definition of the filter associated to a set of positions (see Definition 3.26) and to a set of positions with associated terms (see Definition 3.33).

- Item 2 is a direct consequence of the definition of DN sets of positions (see Definition 3.28) and DN sets of positions with associated terms (see Definition 3.35).

The sets of positions with associated terms of Definition 3.35 were named after "DN" because of the following result.

**Theorem 3.39** *Let $P$ be a binary program and $\tau^+$ be a set of positions with associated terms that is DN for $P$. Then $\Delta[\tau^+]$ is DN for $P$.*

As in the case of sets of positions, the set of DN sets of positions with associated terms of any binary program $P$ is not empty because, by Definition 3.35, $\tau_0^+ := \langle p \mapsto \langle \rangle \mid p \in \Pi \rangle$ is DN for $P$. Moreover, an atom $A$ is $\Delta[\tau_0^+]$-more general than an atom $B$ iff $A$ is more general than $B$. Finally, in Appendix A, we give an incremental algorithm (see Section 4.2) that computes a DN set of positions with associated terms. Its correctness proof is also presented.

## 3.5  Examples

This section presents some examples where we use filters obtained from DN sets of positions and DN sets of positions with associated terms to infer looping queries. As the filters we use in each case are not "empty" (*i.e.* are not obtained from $\tau_0$ or $\tau_0^+$), we are able to compute more looping queries than using the classical subsumption test.

**Example 3.40** *Consider the program* `APPEND` *that we introduced in Example 3.6. Every infinite derivation w.r.t.* `APPEND` *starting from an atomic query only uses the non-unit clause* `C2`. *Therefore, as we aim at inferring looping atomic queries w.r.t.* `APPEND`, *we only focus on* `C2` *in the sequel of this example.*

*As in* `C2` *the body, which is* $append(Xs, Ys, Zs)$, *is more general than the head, which is* $append([X|Xs], Ys, [X|Zs])$, *by Corollary 3.2 we have that the query* $append([X|Xs], Ys, [X|Zs])$ *loops w.r.t.* {`C2`}. *Consequently, by the One Step Lifting Lemma 3.1, each query that is more general than* $append([X|Xs], Ys, [X|Zs])$ *also loops w.r.t.* {`C2`}.

*But we can be more precise than that. According to Definition 3.28, $\tau := \langle\ append \mapsto \{2\}\ \rangle$ is a DN set of positions for* {`C2`}. *The filter associated to $\tau$ (see Definition 3.26) is $\Delta[\tau] := \langle\ append \mapsto \langle 2 \mapsto f_{true} \rangle\ \rangle$. By Theorem 3.30, $\Delta[\tau]$ is a DN filter for* {`C2`}. *Consequently, by Definition 3.18, each query that is $\Delta[\tau]$-more general than* $append([X|Xs], Ys, [X|Zs])$ *loops w.r.t.* {`C2`}. *This means that*

$$\left\{ append(t_1, t_2, t_3) \in TB_{\mathcal{L}} \,\middle|\, \begin{array}{l} t_2 \text{ is any term and} \\ t_1, t_3 \text{ is more general than } [X|Xs], [X|Zs] \end{array} \right\}$$

*is a set of atomic queries that loop w.r.t.* {`C2`}, *hence w.r.t.* `APPEND`. *This set includes the 'well-typed' query* $append(As, [\,], Bs)$. □

**Example 3.41** *Consider the program* `REVERSE` *that was introduced in Example 3.6. As in the example above, in order to infer looping atomic queries w.r.t.* `REVERSE`, *we only focus on the non-unit clauses* `C1` *and* `C3` *in the sequel of this example. More precisely, we process the relation symbols of the program in a bottom-up way, so we start the study with clause* `C3` *and end it with clause* `C1`.

*According to Definition 3.28, $\tau := \langle\ rev \mapsto \{2,3\}\ \rangle$ is a DN set of positions for* {`C3`}. *The filter associated to $\tau$ (see Definition 3.26) is $\Delta[\tau] := \langle\ rev \mapsto \langle 2 \mapsto f_{true},\ 3 \mapsto f_{true} \rangle\ \rangle$. By Theorem 3.30, $\Delta[\tau]$ is DN for* {`C3`}. *As* $rev(Xs, [X|R_0], R)$ *(the body of* `C3`*) is $\Delta[\tau]$-more general than* $rev([X|Xs], R_0, R)$ *(the head of* `C3`*), by Proposition 3.20 we get that* $rev([X|Xs], R_0, R)$ *loops w.r.t.* {`C3`}. *Notice that, unlike the example above, here we do not get this result from Corollary 3.2* as $rev(Xs, [X|R_0], R)$ is not more general than $rev([X|Xs], R_0, R)$. Finally, as $\Delta[\tau]$ is DN for $\{$C3$\}$, by Definition 3.18 we get that each query that is $\Delta[\tau]$-more general than $rev([X|Xs], R_0, R)$ loops w.r.t. $\{$C3$\}$, hence w.r.t. REVERSE. This means that

$$\mathcal{Q} := \left\{ rev(t_1, t_2, t_3) \in TB_{\mathcal{L}} \,\middle|\, \begin{array}{l} t_2 \text{ and } t_3 \text{ are any terms and} \\ t_1 \text{ is more general than } [X|Xs] \end{array} \right\}$$

is a set of atomic queries that loop w.r.t. REVERSE. This set includes the 'well-typed' query $rev(As, [\,], [\,])$.

Now, consider clause C1. As $rev(L, [\,], R)$ (its body) is an element of $\mathcal{Q}$, then $rev(L, [\,], R)$ loops w.r.t. $\{$C3$\}$, hence w.r.t. $\{$C1,C3$\}$. Consequently, by Corollary 3.3, $reverse(L, R)$ (the head of C1) loops w.r.t. $\{$C1,C3$\}$. Let $\tau' := \langle\, rev \mapsto \{2, 3\}, \; reverse \mapsto \{2\} \,\rangle$. By Definition 3.28, $\tau'$ is DN for $\{$C1,C3$\}$, so $\Delta[\tau']$ is DN for $\{$C1,C3$\}$. Consequently, each query that is $\Delta[\tau']$-more general than $reverse(L, R)$ also loops w.r.t. $\{$C1,C3$\}$ hence w.r.t. REVERSE. This means that

$$\{ reverse(X, t) \in TB_{\mathcal{L}} \mid X \text{ is a variable and } t \text{ is any term} \}$$

is a set of atomic queries that loop w.r.t. REVERSE. This set includes the 'well-typed' query $reverse(As, [\,])$. □

**Example 3.42** *Consider the two recursive clauses of the program* MERGE *where we have removed the inequalities:*

```
merge([X|Xs],[Y|Ys],[X|Zs]) :- merge(Xs,[Y|Ys],Zs). % C3
merge([X|Xs],[Y|Ys],[Y|Zs]) :- merge([X|Xs],Ys,Zs). % C4
```

*Every set of positions $\tau$ that is DN for $\{$C3$\}$ is such that $\tau(merge) = \varnothing$ because each argument of the head of C3 is not a variable (see Definition 3.28). Hence, using Proposition 3.20 with a filter obtained from a DN set of positions leads to the same results as using Corollary 3.2: as $merge(Xs, [Y|Ys], Zs)$ is more general than $merge([X|Xs], [Y|Ys], [X|Zs])$, by Corollary 3.2 $merge([X|Xs], [Y|Ys], [X|Zs])$ loops w.r.t. $\{$C3$\}$. So, by the One Step Lifting Lemma 3.1, each query that is more general than $merge([X|Xs], [Y|Ys], [X|Zs])$ also loops w.r.t. $\{$C3$\}$, hence w.r.t. MERGE.*

*But we can be more precise than that. According to Definition 3.35, $\tau^+ := \langle merge \mapsto \langle 2 \mapsto [Y|Ys]\rangle \,\rangle$ is a set of positions with associated terms that is DN for $\{$C3$\}$. Hence, by Theorem 3.39, the associated filter $\Delta[\tau^+]$ (see Definition 3.33) is DN for $\{$C3$\}$. So, by Definition 3.18, each query that is $\Delta[\tau^+]$-more general than $merge([X|Xs], [Y|Ys], [X|Zs])$ loops w.r.t. $\{$C3$\}$. This means that*

$$\left\{ merge(t_1, t_2, t_3) \in TB_{\mathcal{L}} \,\middle|\, \begin{array}{l} t_2 \text{ is any instance of } [Y|Ys] \text{ and} \\ t_1, t_3 \text{ is more general than } [X|Xs], [X|Zs] \end{array} \right\}$$

*is a set of atomic queries that loop w.r.t. MERGE. Notice that this set includes the 'well-typed' query $merge(As, [0], Bs)$. Finally, let us turn to clause C4. Reasoning exactly as above with the set of positions with associated terms $\langle merge \mapsto \langle 1 \mapsto [X|Xs]\rangle \,\rangle$ which is DN for $\{$C4$\}$, we conclude that:*

$$\left\{ merge(t_1, t_2, t_3) \in TB_{\mathcal{L}} \,\middle|\, \begin{array}{l} t_1 \text{ is any instance of } [X|Xs] \text{ and} \\ t_2, t_3 \text{ is more general than } [Y|Ys], [Y|Zs] \end{array} \right\}$$

*is a set of atomic queries that loop w.r.t. MERGE. Notice that this set includes the 'well-typed' query $merge([0], As, Bs)$.* □

# 4 Algorithms

We have designed a set of correct algorithms for full automation of non-termination analysis of logic programs. These algorithms are given in Appendix A with their correctness proofs. In this section, we present the intuitions and conceptual definitions underlying our approach.

### 4.1 Loop Dictionaries

Our technique is based on a data structure called *dictionary* which is a set of pairs $(BinSeq, \tau^+)$ where $BinSeq$ is a finite ordered sequence of binary clauses and $\tau^+$ is a set of positions with associated terms. In the sequel, we use the list notation of Prolog and a special kind of dictionaries that we define as follows.

**Definition 4.1** *(Looping Pair, Loop Dictionary) A pair $(BinSeq, \tau^+)$, where the list $BinSeq$ is a finite ordered sequence of binary clauses and $\tau^+$ is a set of positions with associated terms, is a looping pair if $\tau^+$ is DN for $BinSeq$ and:*

- *either $BinSeq = [H \leftarrow B]$ and $B$ is $\Delta[\tau^+]$-more general than $H$,*

- *or $BinSeq = [H \leftarrow B, H_1 \leftarrow B_1 \mid BinSeq_1]$ and there exists a set of positions with associated terms $\tau_1^+$ such that $([H_1 \leftarrow B_1 \mid BinSeq_1], \tau_1^+)$ is a looping pair and $B$ is $\Delta[\tau_1^+]$-more general than $H_1$.*

*A* loop dictionary *is a finite set of looping pairs.*

**Example 4.2** *The pair $(BinSeq := [H_1 \leftarrow B_1, H_2 \leftarrow B_2, H_3 \leftarrow B_3], \tau_1^+)$ where*

$$H_1 \leftarrow B_1 := r(X) \leftarrow q(X, f(f(X)))$$
$$H_2 \leftarrow B_2 := q(X, f(Y)) \leftarrow p(f(X), a)$$
$$H_3 \leftarrow B_3 := p(f(g(X)), a) \leftarrow p(X, a)$$

*and $\tau_1^+ := \langle p \mapsto \langle 2 \mapsto a \rangle, q \mapsto \langle 2 \mapsto f(X) \rangle \rangle$ is a looping pair:*

- *Let $\tau_3^+ := \langle p \mapsto \langle 2 \mapsto a \rangle \rangle$. Then $\tau_3^+$ is a DN set of positions with associated terms for $[H_3 \leftarrow B_3]$. Moreover, $B_3$ is $\Delta(\tau_3^+)$-more general than $H_3$. Consequently, $([H_3 \leftarrow B_3], \tau_3^+)$ is a looping pair.*

- *Notice that $B_2$ is $\Delta(\tau_3^+)$-more general than $H_3$. Now, let $\tau_2^+ := \tau_1^+$. Then $\tau_2^+$ is DN for $[H_2 \leftarrow B_2, H_3 \leftarrow B_3]$. So, $([H_2 \leftarrow B_2, H_3 \leftarrow B_3], \tau_2^+)$ is a looping pair.*

- *Finally, notice that $B_1$ is $\Delta(\tau_2^+)$-more general than $H_2$. As $\tau_1^+$ is DN for $BinSeq$, we conclude that $(BinSeq, \tau_1^+)$ is a looping pair.* □

A looping pair immediately provides an atomic looping query. It suffices to take the head of the first clause of the binary program of the pair:

**Proposition 4.3** *Let $([H \leftarrow B|BinSeq], \tau^+)$ be a looping pair. Then $H$ loops with respect to $[H \leftarrow B|BinSeq]$.*

*Proof.* By induction on the length of $BinSeq$ using Proposition 3.20, Corollary 3.3 and Theorem 3.39. So, a looping pair denotes a proof outline for establishing that $H$ left loops. Moreover, looping pairs can be built incrementally in a simple way as described below.

### 4.2 Computing a Loop Dictionary

Given a logic program $P$ and a positive integer $max$, the function `infer_loop_dict` from Appendix A first computes $T_P^\beta \uparrow max$ (the first $max$ iterations of the operator $T_P^\beta$), which is a finite subset of $bin\_unf(P)$. Then, using the clauses of $T_P^\beta \uparrow max$, it incrementally builds a loop dictionary $Dict$ as follows.

At start, $Dict$ is set to $\varnothing$. Then, for each clause $H \leftarrow B$ in $T_P^\beta \uparrow max$, the following actions are performed.

- `infer_loop_dict` tries to extract from $H \leftarrow B$ the most simple form of a looping pair: it computes a set of positions with associated terms $\tau^+$ that is DN for $H \leftarrow B$, then it tests if $B$ is $\Delta[\tau^+]$-more general than $H$. If so, the looping pair $([H \leftarrow B], \tau^+)$ is added to $Dict$.

- `infer_loop_dict` tries to combine $H \leftarrow B$ to some looping pairs that have already been added to $Dict$ in order to build other looping pairs. For each $([H_1 \leftarrow B_1 | BinSeq_1], \tau_1^+)$ in $Dict$, if $B$ is $\Delta[\tau_1^+]$-more general than $H_1$, then a set of positions with associated terms $\tau^+$ that is DN for $[H \leftarrow B, H_1 \leftarrow B_1 | BinSeq_1]$ is computed and the looping pair $([H \leftarrow B, H_1 \leftarrow B_1 | BinSeq_1], \tau^+)$ is added to $Dict$.

Notice that in the second step above, we compute $\tau^+$ that is DN for $[H \leftarrow B, H_1 \leftarrow B_1 | BinSeq_1]$. As we already hold $\tau_1^+$ that is DN for $[H_1 \leftarrow B_1 | BinSeq_1]$, it is more interesting, for efficiency reasons, to compute $\tau^+$ from $\tau_1^+$ instead of starting from the ground. Indeed, starting from $\tau_1^+$, one uses the information stored in $\tau_1^+$ about the program $[H_1 \leftarrow B_1 | BinSeq_1]$, which may speed up the computation substantially. This is why we have designed a function `dna` that takes two arguments as input, a binary program $BinProg$ and a set of positions with associated terms $\tau^+$. It computes a set of positions with associated terms that is DN for $BinProg$ and that refines $\tau^+$. On the other hand, the function `unit_loop` calls `dna` with $\tau_{max}^+$ which is the initial set of positions with associated terms defined as follows: $Dom(\tau_{max}^+(p)) = [1, arity(p)]$ for each $p \in \Pi$ and $\tau_{max}^+(p)(i)$ is a variable for each $i \in [1, arity(p)]$.

**Example 4.4** *Consider the program* `APPEND3`

```
append3(Xs,Ys,Zs,Us) :- append(Xs,Ys,Vs), append(Vs,Zs,Us).
```

*augmented with the* `APPEND` *program. The set* $T_{APPEND3}^{\beta} \uparrow 2$ *includes:*

```
append([A|B],C,[A|D]) :- append(B,C,D).        % BC1
append3(A,B,C,D) :- append(A,B,E).             % BC2
append3([],A,B,C) :- append(A,B,C).            % BC3
```

*From clause* `BC1` *we get the looping pair* $(BinSeq_1, \tau_1^+)$ *where*

$$BinSeq_1 = \left[ append([X_1|X_2], X_3, [X_1|X_4]) \leftarrow append(X_2, X_3, X_4) \right]$$

*and* $\tau_1^+(append) = \langle 2 \mapsto X_3 \rangle$. *From this pair and the clause* `BC2`, *we get the looping pair* $(BinSeq_2, \tau_2^+)$ *where:*

$$BinSeq_2 = \left[ \begin{array}{l} append3(X_1, X_2, X_3, X_4) \leftarrow append(X_1, X_2, X_5), \\ append([X_1|X_2], X_3, [X_1|X_4]) \leftarrow append(X_2, X_3, X_4) \end{array} \right]$$

*and* $\tau_2^+(append) = \langle 2 \mapsto X_3 \rangle$ *and* $\tau_2^+(append3) = \langle 2 \mapsto X_3, 3 \mapsto X_3, 4 \mapsto X_4 \rangle$.

*Finally, from* $(BinSeq_1, \tau_1^+)$ *and* `BC3`, *we get the looping pair* $(BinSeq_3, \tau_3^+)$ *where:*

$$BinSeq_3 = \left[ \begin{array}{l} append3([], X_1, X_2, X_3) \leftarrow append(X_1, X_2, X_3), \\ append([X_1|X_2], X_3, [X_1|X_4]) \leftarrow append(X_2, X_3, X_4) \end{array} \right]$$

*and* $\tau_3^+(append) = \langle 2 \mapsto X_3 \rangle$ *and* $\tau_3^+(append3) = \langle 3 \mapsto X_2 \rangle$. $\square$

**Example 4.5** *Consider the program* `PERMUTE`:

```
delete(X,[X|Xs],Xs).
delete(Y,[X|Xs],[X|Ys]) :- delete(Y,Xs,Ys).

permute([],[]).
permute([X|Xs],[Y|Ys]) :- delete(Y,[X|Xs],Zs), permute(Zs,Ys).
```

*The set* $T_{PERMUTE}^{\beta} \uparrow 1$ *includes:*

```
delete(B,[C|D],[C|E]) :- delete(B,D,E).      % BC1
permute([B|C],[D|E]) :- delete(D,[B|C],F).   % BC2
```

*From clause **BC1** we get the looping pair* $(BinSeq_1, \tau_1^+)$ *where*

$$BinSeq_1 = \big[\, delete(X_1, [X_2|X_3], [X_2|X_4]) \leftarrow delete(X_1, X_3, X_4) \,\big]$$

*and* $\tau_1^+(delete) = \langle 1 \mapsto X_1 \rangle$. *From this pair and **BC2**, we get the looping pair* $(BinSeq_2, \tau_2^+)$ *where:*

$$BinSeq_2 = \big[\quad permute([X_1|X_2], [X_3|X_4]) \leftarrow delete(X_3, [X_1|X_2], X_5),$$
$$delete(X_1, [X_2|X_3], [X_2|X_4]) \leftarrow delete(X_1, X_3, X_4) \,\big]$$

*and* $\tau_2^+(delete) = \langle 1 \mapsto X_1 \rangle$ *and* $\tau_2^+(permute) = \langle 2 \mapsto [X_3|X_4] \rangle$. $\qquad\square$

## 4.3 Looping Conditions

One of the main purposes of this article is the inference of classes of atomic queries that left loop w.r.t. a given logic program. Classes of atomic queries we consider are defined by pairs $(A, \tau^+)$ where $A$ is an atom and $\tau^+$ is a set of positions with associated terms. Such a pair denotes the set of queries $A_{\uparrow\tau^+}$, the definition of which is similar to that of the expansion of an atom, see Definition 3.21.

**Definition 4.6** *Let* $A$ *be an atom and* $\tau^+$ *be a set of positions with associated terms. Then* $A_{\uparrow\tau^+}$ *denotes the class of atomic queries defined as:*

$$A_{\uparrow\tau^+} \overset{def}{=} \{A\} \cup \{B \in TB_{\mathcal{L}} \mid B \text{ is } \Delta[\tau^+]\text{-more general than } A\}\,.$$

Once each element of $A_{\uparrow\tau^+}$ left loops w.r.t. a logic program, we get what we call a *looping condition* for that program:

**Definition 4.7** *(Looping Condition) Let* $P$ *be a logic program. A looping condition for* $P$ *is a pair* $(A, \tau^+)$ *such that each element of* $A_{\uparrow\tau^+}$ *left loops w.r.t.* $P$.

The function `infer_loop_cond` takes as arguments a logic program $P$ and a non-negative integer *max*. Calling `infer_loop_dict`$(P, max)$, it first computes a loop dictionary *Dict*. Then, it computes from *Dict* looping conditions for $P$ as follows. The function extracts the pair $(H, \tau^+)$ from each element $([H \leftarrow B|BinSeq], \tau^+)$ of *Dict*. By Proposition 4.3, $H$ loops w.r.t. $[H \leftarrow B|BinSeq]$. As $\tau^+$, hence $\Delta[\tau^+]$, is DN for $[H \leftarrow B|BinSeq]$, by Definition 3.18 each element of $H_{\uparrow\tau^+}$ loops w.r.t. $[H \leftarrow B|BinSeq]$. Finally, as $[H \leftarrow B|BinSeq] \subseteq T_P^\beta \uparrow max \subseteq bin\_unf(P)$, by Theorem 2.1, each element of $H_{\uparrow\tau^+}$ left loops w.r.t. $P$.

**Example 4.8** *(Example 4.4 continued) From each looping pair we have infered, we get the following information.*

- $(append([X_1|X_2], X_3, [X_1|X_4]), \tau_1^+)$ *is a looping condition. So, each query* $append(t_1, t_2, t_3)$, *where* $[X_1|X_2] = t_1\eta$ *and* $[X_1|X_4] = t_3\eta$ *for a substitution* $\eta$ *and* $t_2$ *is an instance of* $X_3$ *(because* $\tau_1^+(append)(2) = X_3$*), left loops w.r.t.* **APPEND3**. *In other words, each query* $append(t_1, t_2, t_3)$, *where* $[X_1|X_2] = t_1\eta$ *and* $[X_1|X_4] = t_3\eta$ *for a substitution* $\eta$ *and* $t_2$ *is any term, left loops w.r.t.* **APPEND3**.

- $(append3(X_1, X_2, X_3, X_4), \tau_2^+)$ *is a looping condition. As we have* $\tau_2^+(append3)(2) = X_2$, $\tau_2^+(append3)(3) = X_3$ *and* $\tau_2^+(append3)(4) = X_4$, *this means that each query of form* $append3(x_1, t_2, t_3, t_4)$, *where* $t_2$, $t_3$ *and* $t_4$ *are any terms, left loops w.r.t.* **APPEND3**.

- $(append3([], X_1, X_2, X_3), \tau_3^+)$ *is a looping condition. So, as* $\tau_3^+(append3)(3) = X_2$, *this means that each query of form* $append3([], X_1, t, X_3)$, *where* $t$ *is any term, left loops w.r.t.* **APPEND3**. $\qquad\square$

**Example 4.9** *(Example 4.5 continued) From each looping pair we have infered, we get the following information.*

- *($delete(X_1, [X_2|X_3], [X_2|X_4]), \tau_1^+$) is a looping condition. As $\tau_1^+(delete)(1) = X_1$, this means that each query of form $delete(t_1, t_2, t_3)$, where $t_1$ is any term and $[X_2|X_3] = t_2\eta$ and $[X_2|X_4] = t_3\eta$ for a substitution $\eta$, left loops w.r.t. PERMUTE.*

- *($permute([X_1|X_2], [X_3|X_4]), \tau_2^+$) is a looping condition. As $\tau_2^+(permute)(2) = [X_3|X_4]$, this means that each query of form $permute(t_1, t_2)$, where $t_1$ is more general than $[X_1|X_2]$ and $t_2$ is any instance of $[X_3|X_4]$, left loops w.r.t. PERMUTE.* □

# 5 An Application: Proving Optimality of Termination Conditions

[26] presents a tool for inferring termination conditions that are expressed as multi-modes, *i.e.* as disjunctions of conjunctions of propositions of form "the $i$-th argument is ground". In this section, we describe an algorithm that attempts proofs of optimality of such conditions using the algorithms for non-termination inference of the previous section.

## 5.1 Optimal Terminating Multi-modes

Let $P$ be a logic program and $p \in \Pi_P$ be a relation symbol, with $arity(p) = n$. First, we describe the language we use for abstracting sets of atomic queries:

**Definition 5.1** *(Mode) A mode $m_p$ for $p$ is a subset of $[1, n]$, and denotes the following set of atomic goals: $[m_p] = \{p(t_1, \ldots, t_n) \in TB_{\mathcal{L}} \mid \forall i \in m_p \; Var(t_i) = \varnothing\}$. The set of all modes for $p$, i.e. $2^{[1,n]}$, is denoted $modes(p)$.*

Note that if $m_p = \varnothing$ then $[m_p] = \{p(t_1, \ldots, t_n) \in TB_{\mathcal{L}}\}$. Since a logic procedure may have multiple uses, we generalize:

**Definition 5.2** *(Multi-mode) A multi-mode $M_p$ for $p$ is a finite set of modes for $p$ and denotes the following set of atomic queries: $[M_p] = \cup_{m \in M_p}[m]$.*

Note that if $M_p = \varnothing$, then $[M_p] = \varnothing$. Now we can define what we mean by terminating and looping multi-modes:

**Definition 5.3** *(Terminating mode, terminating multi-mode) A terminating mode $m_p$ for $p$ is a mode for $p$ such that any query in $[m_p]$ left terminates w.r.t. $P$. A terminating multi-mode $TM_p$ for $p$ is a finite set of terminating modes for $p$.*

**Definition 5.4** *(Looping mode, looping multi-mode) A looping mode $m_p$ for $p$ is a mode for $p$ such that there exists a query in $[m_p]$ which left loops w.r.t. $P$. A looping multi-mode $LM_p$ for $p$ is a finite set of looping modes for $p$.*

As left termination is instantiation-closed, any mode that is "below" (less general than) a terminating mode is also a terminating mode. Similarly, as left looping is generalization-closed, any mode that is "above" (more general than) a looping mode is also a looping mode. Let us be more precise:

**Definition 5.5** *(Less_general, more_general) Let $M_p$ be a multi-mode for the relation symbol $p$. We set:*

$$less\_general(M_p) = \{m \in modes(p) \mid \exists m' \in M_p \; [m] \subseteq [m']\}$$
$$more\_general(M_p) = \{m \in modes(p) \mid \exists m' \in M_p \; [m'] \subseteq [m]\}$$

**in**:  $L$: a finite set of looping conditions

  $p$: a predicate symbol

**out**:  a looping multi-mode for $p$

1:  $LM_p := \varnothing$

2:  **for each**  $(p(t_1, \ldots, t_n), \tau^+) \in L$  **do**

3:    $m_p := Dom(\tau^+(p)) \ \cup \ \{i \in [1, n] \mid Var(t_i) = \varnothing\}$

4:    $LM_p := LM_p \cup \{m_p\}$

5:  **return**  $LM_p$

Figure 1:

We are now equipped to present a definition of optimality for terminating multi-modes:

**Definition 5.6**  *(Optimal terminating multi-mode) A terminating multi-mode $TM_p$ for $p$ is optimal if there exists a looping multi-mode $LM_p$ verifying:*

$$modes(p) = less\_general(TM_p) \cup more\_general(LM_p)$$

Otherwise stated, given a terminating multi-mode $TM_p$, if each mode which is not less general than a mode of $TM_p$ is a looping mode, then $TM_p$ *characterizes* the operational behavior of $p$ w.r.t. left termination and our language for defining sets of queries.

**Example 5.7**  *Consider the program `APPEND`. A well-known terminating multi-mode is the set $TM_{append} = \{\{1\}, \{3\}\}$. Indeed, any query of the form `append(t,Ys,Zs)` or `append(Xs,Ys,t)`, where $t$ is a ground term (i.e. such that $Var(t) = \varnothing$), left terminates. We have:*

$$less\_general(TM_{append}) = \{\{1\}, \{3\}, \{1, 2\}, \{1, 3\}, \{2, 3\}, \{1, 2, 3\}\}$$

*On the other hand, `append(Xs,[],Zs)` left loops. Hence $LM_{append} = \{\{2\}\}$ is a looping condition and $more\_general(LM_{append}) = \{\varnothing, \{2\}\}$.*

*Since $modes(append) = less\_general(TM_{append}) \cup more\_general(LM_{append})$, we conclude that the terminating multi-mode $TM_{append}$ is optimal.*  □

## 5.2  Algorithms

Suppose we hold a finite set $L$ of looping conditions for $P$. Then, each element $(p(t_1, \ldots, t_n), \tau^+)$ of $L$ provides an obvious looping mode for $p$: it suffices to take $\{i \in [1, n] \mid Var(t_i) = \varnothing\}$. But actually, we can extract more information from $L$. Let $p(t'_1, \ldots, t'_n)$ be an atom such that:

- for each $\langle i \mapsto u_i \rangle \in \tau^+(p)$, $t'_i$ is a ground instance of $u_i$,

- for each $i$ in $[1, n] \setminus Dom(\tau^+(p))$, $t'_i = t_i$.

Then, $p(t'_1, \ldots, t'_n)$ belongs to $p(t_1, \ldots, t_n)_{\uparrow \tau^+}$, hence it left loops w.r.t.  $P$.  Consequently, we have that $Dom(\tau^+(p)) \ \cup \ \{i \in [1, n] \mid Var(t_i) = \varnothing\}$ is a looping mode for $p$. The function `looping_modes` of Fig. 1 is an application of these remarks.

Now we have the essential material for the design of a tool that attempts proofs of optimality of left terminating multi-modes computed by a termination inference tool as e.g.  cTI [26] or TerminWeb [17]. For each pair $(p, \varnothing)$ in the set the function `optimal_tc` of Fig. 2 returns, we can conclude that the corresponding $TM_p$ is *the* optimal terminating multi-mode which characterizes the operational behavior of $p$ with respect to $\mathcal{L}_{term}$.

```
optimal_tc(P, max, {TM_p}_{p∈Π_P}):
```

| | |
|---|---|
| **in:** | $P$: a logic program |
| | $max$: a non-negative integer |
| | $\{TM_p\}_{p∈Π_P}$: a finite set of terminating multi-modes |
| **out:** | a finite set of pairs $(p, M_p)$ such that $p ∈ Π_P$ and |
| | $M_p$ is a multi-mode for $p$ with no information w.r.t. its left behaviour |
| **note:** | if for each $p ∈ Π_P$, $M_p = \varnothing$, then $\{TM_p\}_{p∈Π_P}$ is optimal |

1: $Res := \varnothing$
2: $L := \texttt{infer\_loop\_cond}(P, max)$
3: **for each** $p ∈ Π_P$ **do**
4: $\quad LM_p := \texttt{looping\_modes}(L, p)$
5: $\quad M_p := \texttt{modes}(p) \setminus (\texttt{less\_general}(TM_p) ∪ \texttt{more\_general}(LM_p))$
6: $\quad Res := Res ∪ \{(p, M_p)\}$
7: **return** $Res$

Figure 2:

**Example 5.8** *(Example 4.8 continued) We apply our algorithm to the program* **APPEND3** *of Example 4.4. We get that*

$$L := \{ \quad (append([X_1|X_2], X_3, [X_1|X_4]), \quad \tau_1^+),$$
$$(append3(X_1, X_2, X_3, X_4), \quad\quad\quad \tau_2^+),$$
$$(append3([], X_1, X_2, X_3), \quad\quad\quad \tau_3^+) \quad \}$$

*is a finite set of looping conditions for* **APPEND3** *(see Example 4.8) with*

$$Dom(\tau_1^+(append)) = \{2\}$$
$$Dom(\tau_2^+(append3)) = \{2, 3, 4\}$$
$$Dom(\tau_3^+(append3)) = \{3\}$$

*So, for append we have:*

$$LM_{append} := \texttt{looping\_modes}(L, append) = \{\{2\}\}$$
$$\texttt{more\_general}(LM_{append}) = \{\varnothing, \{2\}\}$$
$$TM_{append} = \{\{1\}, \{3\}\}$$
$$\texttt{less\_general}(TM_{append}) = \{\{1\}, \{3\}, \{1, 2\}, \{1, 3\}, \{2, 3\}, \{1, 2, 3\}\}$$
$$M_{append} = \{\}$$

*For append3, we get:*

- *the looping mode $\{2, 3, 4\}$ from $(append3(X_1, X_2, X_3, X_4), \tau_2^+)$ and*

- *the looping mode $m_p := \{1, 3\}$ from $(append3([], X_1, X_2, X_3), \tau_3^+)$ (notice that $3 ∈ m_p$ because $Dom(\tau_3^+(append3)) = \{3\}$ and $1 ∈ m_p$ because of constant $[]$ which is the first argument of $append3([], X_1, X_2, X_3)$.)*

*So, we have:*

$$
\begin{aligned}
LM_{append3} &:= \texttt{looping\_modes}(L, append3) = \{\{2,3,4\},\{1,3\}\} \\
\texttt{more\_general}(LM_{append3}) &= \{\varnothing, \{1\},\{2\},\{3\},\{4\},\{1,3\},\{2,3\},\{2,4\}, \\
&\qquad \{3,4\},\{2,3,4\}\} \\
TM_{append3} &= \{\{1,2\},\{1,4\}\} \\
\texttt{less\_general}(TM_{append3}) &= \{\{1,2\},\{1,4\},\{1,2,3\},\{1,2,4\},\{1,3,4\}, \\
&\qquad \{1,2,3,4\}\} \\
M_{append3} &= \{\}
\end{aligned}
$$

*Hence in both cases, we have characterized the left behaviour of the predicates by using two complementary tools.* □

## 5.3 An Experimental Evaluation

We have implemented[1] the algorithms presented in Sections 4 and 5.2. The binary unfoldings algorithm is derived from the one described in [7], where we added time stamps to precisely control what is computed at each iteration. Looping modes are computed starting from the leaves of the call graph then moving up to its roots. The cTI termination inference tool[2] is detailed in [26, 24]. Here is the configuration we used for this experiment: Intel 686, 2.4GHz, 512Mb, Linux 2.4, SICStus Prolog 3.10.1, 24.8 MLips. Timings in seconds are average over 10 runs.

First we have applied them on some small programs from standard benchmarks of the termination analysis literature [30, 2, 9] (predefined predicates were erased). The column *opt?* of Table 1 indicates whether the result of cTI (see [26]) is proved optimal ($\checkmark$) or not ($?$). The column *max* gives the least non-negative integer implying optimality or the least non-negative integer $n$ where it seems we get the most precise information from non-termination inference (*i.e.* for $n$ and $n+1$, the analyser delivers the same results). Then timings in seconds (t[s]) appear, followed by a pointer to a comment to the notes below.

**Notes:**

1. The predicate `fold/3` is defined by:

   ```
   fold(X,[],X).
   fold(X,[Y|Ys],Z) :- op2(X,Y,V), fold(V,Ys,Z).
   ```

   When the predicate `op2/3` is defined by the fact `op2(A,B,C)`, the result of cTI is optimal. When the predicate `op2/3` is defined by the fact `op2(a,b,c)`, no looping mode is found and the result of cTI is indeed sub-optimal as the query `fold(X,Y,Z)` terminates.

2. Termination proofs for `mergesort` require the *list-size* norm, while cTI applies the *term-size* norm.

3. The result of cTI is not optimal. The analyzed program:

   ```
   p(A,B) :- q(A,C),p(C,B).
   p(A,A).
   q(a,b).
   ```

   has finite binary unfoldings because there is no function symbol. Hence its termination is decidable (see [7]). This could be easily detected at analyze time. We notice that no looping mode is found. But as any constant is mapped to 0 by the *term-size* norm, the modes $modes(p)$ remain undecided for cTI while they all terminate.

---

[1] Available from `http://www.univ-reunion.fr/~gcc`
[2] Available from `http://www.cs.unipr.it/cTI`

Table 1: Some De Schreye's, Apt's, and Plümer's programs.

| program | top-level predicate | cTI | | Optimal | | | cf. |
| | | term-cond | t[s] | opt? | max | t[s] | |
|---|---|---|---|---|---|---|---|
| permute | permute(X,Y) | $X$ | 0.01 | ✓ | 1 | 0.01 | |
| duplicate | duplicate(X,Y) | $X \vee Y$ | 0.01 | ✓ | 1 | 0.01 | |
| sum | sum(X,Y,Z) | $X \vee Y \vee Z$ | 0.01 | ✓ | 1 | 0.01 | |
| merge | merge(X,Y,Z) | $(X \wedge Y) \vee Z$ | 0.02 | ✓ | 1 | 0.01 | |
| dis-con | dis(X) | $X$ | 0.02 | ✓ | 2 | 0.01 | |
| reverse | reverse(X,Y,Z) | $X$ | 0.02 | ✓ | 1 | 0.01 | |
| append | append(X,Y,Z) | $X \vee Z$ | 0.01 | ✓ | 1 | 0.01 | |
| list | list(X) | $X$ | 0.01 | ✓ | 1 | 0.01 | |
| fold | fold(X,Y,Z) | $Y$ | 0.01 | ? | 2 | 0.01 | note 1 |
| lte | goal | 1 | 0.01 | ✓ | 1 | 0.01 | |
| map | map(X,Y) | $X \vee Y$ | 0.01 | ✓ | 2 | 0.01 | |
| member | member(X,Y) | $Y$ | 0.01 | ✓ | 1 | 0.01 | |
| mergesort | mergesort(X,Y) | $X$ | 0.04 | ? | 2 | 0.01 | note 2 |
| mergesort_ap | mergesort_ap(X,Y,Z) | $Z$ | 0.08 | ? | 2 | 0.02 | |
| naive_rev | naive_rev(X,Y) | $X$ | 0.02 | ✓ | 1 | 0.01 | |
| ordered | ordered(X) | $X$ | 0.01 | ✓ | 1 | 0.01 | |
| overlap | overlap(X,Y) | $X \wedge Y$ | 0.01 | ✓ | 2 | 0.01 | |
| permutation | permutation(X,Y) | $X$ | 0.03 | ✓ | 1 | 0.01 | |
| quicksort | quicksort(X,Y) | $X$ | 0.05 | ✓ | 1 | 0.01 | |
| select | select(X,Y,Z) | $Y \vee Z$ | 0.01 | ✓ | 1 | 0.01 | |
| subset | subset(X,Y) | $X \wedge Y$ | 0.01 | ✓ | 2 | 0.01 | |
| sum_peano | sum(X,Y,Z) | $Y \vee Z$ | 0.01 | ✓ | 1 | 0.01 | |
| pl2.3.1 | p(X,Y) | 0 | 0.01 | ? | 1 | 0.01 | note 3 |
| pl3.5.6 | p(X) | $X$ | 0.01 | ✓ | 2 | 0.01 | |
| pl4.4.6a | perm(X,Y) | $X$ | 0.02 | ✓ | 1 | 0.01 | |
| pl4.5.2 | s(X,Y) | 0 | 0.03 | ✓ | 1 | 0.01 | |
| pl4.5.3a | p(X) | 0 | 0.01 | ✓ | 1 | 0.01 | |
| pl5.2.2 | turing(X,Y,Z,T) | 0 | 0.08 | ? | 2 | 0.03 | note 4 |
| pl7.2.9 | mult(X,Y,Z) | $X \wedge Y$ | 0.02 | ✓ | 4 | 0.03 | note 5 |
| pl7.6.2a | reach(X,Y,Z) | 0 | 0.02 | ? | 1 | 0.01 | note 6 |
| pl7.6.2b | reach(X,Y,Z,T) | 0 | 0.02 | ? | 1 | 0.01 | |
| pl7.6.2c | reach(X,Y,Z,T) | $Z \wedge T$ | 0.02 | ? | 2 | 0.02 | |
| pl8.3.1a | minsort(X,Y) | $X$ | 0.03 | ✓ | 2 | 0.02 | |
| pl8.4.1 | even(X) | $X$ | 0.02 | ✓ | 2 | 0.01 | |
| pl8.4.2 | e(X,Y) | $X$ | 0.05 | ✓ | 3 | 0.04 | |

Table 2: Some middle-sized programs.

| program | | cTI | | Optimal | | | | | |
|---------|---------|-----|------|---------|------|---------|------|---------|------|
| | | | | max=1 | | max=2 | | max=3 | |
| name | clauses | Q% | t[s] | Opt% | t[s] | Opt% | t[s] | Opt% | t[s] |
| `ann` | 177 | 48 | 1.00 | 46 | 0.14 | 68 | 1.34 | 74 | 32.4 |
| `bid` | 50 | 100 | 0.14 | 55 | 0.02 | 90 | 0.08 | 95 | 0.50 |
| `boyer` | 136 | 84 | 0.30 | 80 | 0.03 | 96 | 0.22 | 100 | 3.66 |
| `browse` | 30 | 53 | 0.26 | 46 | 0.03 | 80 | 0.18 | 100 | 6.05 |
| `credit` | 57 | 100 | 0.11 | 91 | 0.02 | 95 | 0.11 | 100 | 4.46 |
| `peephole` | 134 | 88 | 1.08 | 23 | 0.06 | 70 | 3.62 | 70 | 406 |
| `plan` | 29 | 100 | 0.11 | 68 | 0.02 | 81 | 0.09 | 81 | 0.37 |
| `qplan` | 148 | 61 | 1.13 | 50 | 0.11 | 79 | 1.60 | 81 | 1911 |
| `rdtok` | 55 | 44 | 0.65 | 44 | 0.11 | 88 | 40.2 | ? | > 3600 |
| `read` | 88 | 52 | 1.72 | 39 | 0.04 | 47 | 0.80 | 47 | 10.9 |
| `warplan` | 101 | 32 | 0.49 | 37 | 0.07 | 83 | 0.99 | 91 | 21.5 |

4. The analyzed program (from [30], p. 64) simulates a Turing machine. The result of cTI is optimal.

5. With respect to the program:

```
mult(0,A,0).
mult(s(A),B,C) :- mult(A,B,D),add(D,B,C).

add(0,A,A).
add(s(A),B,s(C)) :- add(A,B,C).
```

the query `mult(s(s(0)),A,B)` is automatically detected as looping, although `mult(0,A,B)` and `mult(s(0),A,B)` do terminate.

6. These three programs propose various definitions of the reachability relation between two nodes in a list of edges. For the first and the third definition, cTI is indeed optimal. For the second one, cTI is not optimal.

Next, we have applied the couple of analyzers to some middle-sized Prolog programs, see Table 2. Again, predefined predicates were all erased, while they are usually taken into account for cTI which of course improves the analysis. In other words, we only consider the logic programming skeleton of each program. The first two columns give the name of the file and its size (number of clauses). The fourth column indicates the running time (in seconds) of the termination analysis, while the third column is the ratio of predicates for which a non-false termination condition is computed over the total number of predicates defined in the program. For instance, cTI is able to show that there is at least one terminating mode for 48% of the predicates defined in the program `ann`. We ran the non-termination analyzer with $1 \leq max \leq 3$ iterations. For each value of $max$, we give the running time (in seconds) and the ratio of predicates for which looping modes complement terminating modes. For example, with respect to the program `ann`, for $max = 2$ we get the full *complete* mode termination behavior of 68% of all the defined predicates.

We note that when we increase $max$, we obtain better results but the running times also increase, which is fairly obvious. For $max = 3$, we get good to optimal results but the binary unfoldings approach reveals its potentially explosive nature: we aborted the analysis of `rdtok` after one hour of computation.

In conclusion, from such a naive implementation, we were rather surprised by the quality of the combined analysis. Adopting some more clever implementation schemes, for instance computing the binary unfoldings in a demand driven fashion, could be investigated to improve the running times.

# 6  Related Works

Some extensions of the Lifting Theorem with respect to infinite derivations are presented in [18], where the authors study numerous properties of finite failure. The non-ground finite failure set of a logic program is defined as the set of possibly non-ground atoms which admit a fair finitely failed SLD-tree w.r.t. the program. This denotation is shown correct in the following sense. If two programs have the same non-ground finite failure set, then any ground or non-ground goal which finitely fails w.r.t. one program also finitely fails w.r.t. the other. Such a property is false when we consider the standard ground finite failure set. The proof of correctness of the non-ground finite failure semantics relies on the following result. First, a derivation is called non-perpetual if it is a fair infinite derivation and there exists a finite depth from which unfolding does not instantiate the original goal any more. Then the authors define the definite answer goal of a non-perpetual derivation as the maximal instantiation of the original goal. A crucial lemma states that any instance of the definite answer goal admits a similar non-perpetual derivation. Compared to our work, note that we do not need fairness as an hypothesis for our results. On the other hand, investigating the relationships between non-ground arguments of the definite answer and neutral arguments is an interesting problem.

In [35], the authors present a dynamic approach to characterize (in the form of a necessary and sufficient condition) termination of general logic programs. Their technique employs some key dynamic features of an infinite generalized SLDNF-derivation, such as repetition of selected subgoals and recursive increase in term size.

Loop checking in logic programming is also a subject related to our work. In this area, [5] sets up some solid foundations. A loop check is a device to prune derivations when it seems appropriate. A loop checker is defined as *sound* if no solution is lost. It is *complete* if all infinite derivations are pruned. A complete loop check may also prune finite derivations. The authors show that even for function-free programs (also known as Datalog programs), sound and complete loop checks are out of reach. Completeness is shown only for some restricted classes of function-free programs.

We now review loop checking in more details. To our best knowledge, among all existing loop checking mechanisms only OS-check [32], EVA-check [34] and VAF-check [36] are suitable for logic programs with function symbols. They rely on a structural characteristic of infinite SLD-derivations, namely, the growth of the size of some generated subgoals. This is what the following theorem states.

**Theorem 6.1** *Consider an infinite SLD-derivation $\xi$ where the leftmost selection rule is used. Then there are infinitely many queries $Q_{i_1}$, $Q_{i_2}$, ... (with $i_1 < i_2 < \dots$) in $\xi$ such that for any $j \geq 1$, the selected atom $A_{i_j}$ of $Q_{i_j}$ is an ancestor of the selected atom $A_{i_{j+1}}$ of $Q_{i_{j+1}}$ and $size(A_{i_{j+1}}) \geq size(A_{i_j})$.*

Here, *size* is a given function that maps an atom to its size which is defined in terms of the number of symbols appearing in the atom. As this theorem does not provide any sufficient condition to detect infinite SLD-derivations, the three loop checking mechanisms mentioned above may detect finite derivations as infinite. However, these mechanisms are *complete w.r.t. the leftmost selection rule i.e.* they detect all infinite loops when the leftmost selection rule is used.

OS-check (for OverSize loop check) was first introduced by Shalin [31, 32] and was then formalized by Bol [4]. It is based on a function *size* that can have one of the three following definitions: for any atoms $A$ and $B$, either $size(A) = size(B)$, either $size(A)$ (resp. $size(B)$) is the count of symbols appearing in $A$ (resp. $B$), either $size(A) \leq size(B)$ if for each $i$, the count of symbols of the $i$-th argument of $A$ is smaller than or equal to that of the $i$-th argument of $B$. OS-check says that an SLD-derivation may be infinite if it generates an atomic subgoal $A$ that is *oversized, i.e.* that has ancestor subgoals which have the same predicate symbol as $A$ and whose size is smaller than or equal to that of $A$.

EVA-check (for Extented Variant Atoms loop check) was introduced by Shen [34]. It is based on the notion of *generalized variants* (if $G_i$ and $G_j$ $(i < j)$ are two goals in an SLD-derivation, an atom $A$ in $G_j$ is a generalized variant of an atom $A'$ in $G_i$ if $A$ is a variant of $A'$ except for some arguments whose size increases from $A'$ to $A$ via a set of recursive clauses.) EVA-check says that an SLD-derivation may be infinite if it generates an atomic subgoal $A$ that is a generalized variant of some of its ancestor $A'$. Here the size function that is used applies to predicate arguments, *i.e.* to terms, and it is fixed: it is defined as the the count of symbols that appear in the terms. EVA-check is more reliable than OS-check because it is less likely to mis-identify infinite loops [34]. This is mainly due to the fact that, unlike OS-check, EVA-check refers to the informative internal structure of subgoals.

VAF-check (for Variant Atoms loop check for logic programs with Functions) was proposed by Shen *et al.* [36]. It is based on the notion of *expanded variants*. An atom $A$ is an expanded variant of an atom $A'$ if, after variable renaming, $A$ becomes $B$ that is the same as $A'$ except that there may be some terms at certain positions in $A'$, each $A'[i] \ldots [k]$ of which grows in $B$ into a function $B[i] \ldots [k] = f(\ldots, A'[i] \ldots [k], \ldots)$ (here, we use $A'[i] \ldots [k]$ (resp. $B[i] \ldots [k]$) to refer to the $k$-th argument of $\ldots$ of the $i$-th argument of $A'$ (resp. $B$)). VAF-check says that an SLD-derivation may be infinite if it generates an atomic subgoal $A$ that is an expanded variant of some of its ancestor $A'$. VAF-check is as reliable as and more efficient than EVA-check [36].

The main difference with our work is that we want to infer atomic queries which are guaranteed to be left looping. Hence, we consider *sufficient* conditions for looping, in contrast to the above mentioned methods which consider *necessary* conditions. Our technique returns a set of queries for which it has pinpointed *one* infinite derivation. Hence, we are not interested in soundness as we do not care of finite derivations, nor in completeness, as the existence of just one infinite derivation suffices. Of course, using the $\Delta$-subsumption test as a loop checker leads to a device that is neither sound (since $\Delta$-subsumption is a particular case of subsumption) nor complete (since the $\Delta$-subsumption test provides a sufficient but not necessary condition). This is illustrated by the following example.

**Example 6.2** *Let $c := p(X, X) \leftarrow p(f(X), f(X))$. As the arguments of the head of $c$ have one common variable $X$, every set of positions with associated terms $\tau^+$ that is DN for $\{c\}$ is such that the domain of $\tau^+(p)$ is empty (see **(DN1)** in Definition 3.35). Notice that from the head $p(X, X)$ of $c$ we get*

$$p(X, X) \underset{c}{\Longrightarrow} p(f(X), f(X)) \underset{c}{\Longrightarrow} \cdots \underset{c}{\Longrightarrow} p(f^n(X), f^n(X)) \underset{c}{\Longrightarrow} \cdots$$

*As the arguments of $p$ grow from step to step, there cannot be any query in the derivation that is $\Delta[\tau^+]$-more general than one of its ancestors. Consequently, we can not conclude that $p(X, X)$ left loops w.r.t $\{c\}$.* $\square$

On the other hand, using loop checking approaches to infer classes of atomic left looping queries is not satisfactory because, as we said above, non-looping queries may be mis-identified as looping.

**Example 6.3** *We cannot replace, in Corollary 3.2, the subsumption test by the expanded variant test used in VAF-check because, for instance, in the clause $c := p(a) \leftarrow p(f(a))$, we have: $p(f(a))$ is an expanded variant of $p(a)$, but $p(a)$ does not loop w.r.t. $c$.*

Finally, [10] is also related to our study. In this paper, the authors describe an algorithm for detecting non-terminating queries to clauses of the type $p(\cdots) \leftarrow p(\cdots)$. The algorithm is able to check if such a given clause has no non-terminating queries or has a query which either loops or fails due to occur check. Moreover, given a *linear* atomic goal (*i.e.* a goal where all variable occurs at most once), the algorithm is able to check if the goal loops or not w.r.t. the clause. The technique of the algorithm is based on directed weighted graphs [14] and on a necessary and sufficient condition for the existence of non-terminating queries to clauses of the type $p(\cdots) \leftarrow p(\cdots)$. This condition is proved in [8] and is expressed in terms of rational trees.

# 7 Conclusion

We have presented a extension of the subsumption test which allows to disregard some arguments, termed neutral arguments, while checking for subsumption. We have proposed two syntactic criteria for statically identifying neutral arguments. From these results, in the second part of this report we have described algorithms for automating non-termination analysis of logic programs, together with correctness proofs. Finally, we have applied these techniques to check the optimality of termination conditions for logic programs.

This paper leaves numerous questions open. For instance, it might be interesting to try to generalize this approach to constraint logic programming [19]. Can we obtain higher level proofs compared to those we give? Can we propose more abstract criteria for identifying neutral arguments? A first step in this direction is presented in [29]. Also, our work aims at inferring classes of atomic left looping queries, using a bottom-up point of view. Experimental data show that it may sometimes lead to prohibitive time/space costs. How can we generate only the useful binary clauses without fully computing the iterations of this $T_P$-like operator? Or can we adapt our algorithms towards a more efficient correct top-down approach for checking non-termination?

## Acknowledgments

We thank Ulrich Neumerkel for numerous discussions on this topic, Roberto Bagnara and anonymous referees for interesting suggestions.

# A Algorithms

First, we define a pre-order relation over sets of positions with associated terms. Such a relation is useful for the design of the algorithms that we present in the sequel of this section.

**Definition A.1** ($\preccurlyeq$ and $\tau_{max}^+$)

- $\tau_1^+ \preccurlyeq \tau_2^+$ if for each $p \in \Pi$, $Dom(\tau_1^+(p)) \subseteq Dom(\tau_2^+(p))$ and for each $i \in Dom(\tau_1^+(p))$, $\tau_2^+(p)(i)$ is more general than $\tau_1^+(p)(i)$.

- $\tau_{max}^+$ denotes a set of positions with associated terms s.t. $Dom(\tau_{max}^+(p)) = [1, arity(p)]$ for each $p \in \Pi$ and $\tau_{max}^+(p)(i)$ is a variable for each $i \in [1, arity(p)]$.

## A.1 DN Sets of Positions with Associated Terms for Binary Programs

We present below an algorithm for computing DN sets of positions with associated terms.

---
$\mathtt{dna}\big(BinProg, \tau_1^+\big)$:

   **in:**   $BinProg$: a finite set of binary clauses

           $\tau_1^+$: a set of positions with associated terms

 **out:**   $\tau_2^+$ s.t. $\tau_2^+ \preccurlyeq \tau_1^+$ and $\tau_2^+$ is DN for $BinProg$

   1:   $\tau_2^+ := \tau_1^+$

   2:   $\tau_2^+ := \mathtt{satisfy\_DN1}(BinProg, \tau_2^+)$

   3:   $\tau_2^+ := \mathtt{satisfy\_DN2}(BinProg, \tau_2^+)$

   4:   $\tau_2^+ := \mathtt{satisfy\_DN3}(BinProg, \tau_2^+)$

   5:   **while** $\mathtt{satisfy\_DN4}(BinProg, \tau_2^+) \neq \tau_2^+$ **do**

   6:       $\tau_2^+ := \mathtt{satisfy\_DN4}(BinProg, \tau_2^+)$

   7:   **return** $\tau_2^+$

---

The algorithm $\mathtt{dna}$ calls four auxiliary functions that correspond to conditions **(DN1)**, **(DN2)** **(DN3)** and **(DN4)** in the definition of a DN set of positions with associated terms (see Definition 3.35). These functions are detailed below.

After $\tau_2^+ := \mathtt{satisfy\_DN1}(BinProg, \tau_2^+)$ at line 2 of $\mathtt{dna}$, $\tau_2^+$ satisfies item **(DN1)** of Definition 3.35.

---
$\mathtt{satisfy\_DN1}\big(BinProg, \tau_1^+\big)$:

   1:   $\tau_2^+ := \tau_1^+$

   2:   **for each** $p(s_1, \ldots, s_n) \leftarrow B \in BinProg$ **do**

   3:       $E := \{i \in [1, n] \mid Var(s_i) \cap Var(\{s_j \mid j \neq i\}) = \varnothing\}$

   4:       $\tau_2^+(p) := \tau_2^+(p)|(Dom(\tau_2^+(p)) \cap E)$

   5:   **return** $\tau_2^+$

---

After $\tau_2^+ := \mathtt{satisfy\_DN2}(BinProg, \tau_2^+)$ at line 3 of $\mathtt{dna}$, $\tau_2^+$ satisfies item **(DN2)** of Definition 3.35. Notice that $\mathtt{less\_general}$ at line 5 of $\mathtt{satisfy\_DN2}$ is a function that returns the less general term of two given terms; if none of the given terms is less general than the other, then this function returns *undefined*.

---

$\texttt{satisfy\_DN2}\big(BinProg,\ \tau_1^+\big):$

1:   $\tau_2^+ := \tau_1^+$

2:   **for each** $p(s_1,\ldots,s_n) \leftarrow B \in BinProg$ **do**

3:     $F := \varnothing$

4:     **for each** $i \in Dom(\tau_2^+(p))$ **do**

5:       $u_i := \texttt{less\_general}(s_i, \tau_2^+(p)(i))$

6:       **if** $u_i = undefined$ **then** $F := F \cup \{i\}$

7:       **else** $\tau_2^+(p)(i) := u_i$

8:     $\tau_2^+(p) := \tau_2^+(p)|(Dom(\tau_2^+(p)) \setminus F)$

9:   **return** $\tau_2^+$

---

After $\tau_2^+ := \texttt{satisfy\_DN3}(BinProg, \tau_2^+)$ at line 4 of $\texttt{dna}$, $\tau_2^+$ satisfies item **(DN3)** of Definition 3.35. The function $\texttt{satisfy\_DN3}$ is detailed below.

---

$\texttt{satisfy\_DN3}\big(BinProg,\ \tau_1^+\big):$

1:   $\tau_2^+ := \tau_1^+$

2:   **for each** $H \leftarrow q(t_1,\ldots,t_m) \in BinProg$ **do**

3:     $F := \varnothing$

4:     **for each** $i \in Dom(\tau_2^+(q))$ **do**

5:       **if** $\tau_2^+(q)(i)$ is not more general than $t_i$ **then** $F := F \cup \{i\}$

6:     $\tau_2^+(q) := \tau_2^+(q)|(Dom(\tau_2^+(q)) \setminus F)$

7:   **return** $\tau_2^+$

---

Finally, the function $\texttt{satisfy\_DN4}$ is defined as follows. After line 6 of $\texttt{dna}$, the set of positions with associated terms $\tau_2^+$ satisfies item **(DN4)** of Definition 3.35.

---

$\texttt{satisfy\_DN4}\big(BinProg,\ \tau_1^+\big):$

1:   $\tau_2^+ := \tau_1^+$

2:   **for each** $p(s_1,\ldots,s_n) \leftarrow q(t_1,\ldots,t_m) \in BinProg$ **do**

3:     $F := \varnothing$

4:     **for each** $i \in Dom(\tau_2^+(p))$ **do**

5:       **for each** $j \in [1, m] \setminus Dom(\tau_2^+(q))$ **do**

6:         **if** $Var(s_i) \cap Var(t_j) \neq \varnothing$ **then** $F := F \cup \{i\}$

7:     $\tau_2^+(p) := \tau_2^+(p)|(Dom(\tau_2^+(p)) \setminus F)$

8:   **return** $\tau_2^+$

---

**Proposition A.2** *(Correctness of $\texttt{dna}$) Let $BinProg$ be a binary program and $\tau_1^+$ be a set of positions with associated terms.*

1. *$\texttt{satisfy\_DN1}(BinProg, \tau_1^+)$, ..., $\texttt{satisfy\_DN4}(BinProg, \tau_1^+)$ terminate;*

2. *$\texttt{satisfy\_DN1}(BinProg, \tau_1^+) \preccurlyeq \tau_1^+$, ..., $\texttt{satisfy\_DN4}(BinProg, \tau_1^+) \preccurlyeq \tau_1^+$;*

3. *$\texttt{dna}(BinProg, \tau_1^+)$ terminates;*

4. *$\texttt{dna}(BinProg, \tau_1^+) \preccurlyeq \tau_1^+$ and $\texttt{dna}(BinProg, \tau_1^+)$ is a set of positions with associated terms that is DN for $BinProg$.*

*Proof.* We have:

1. $\texttt{satisfy\_DN1}(BinProg, \tau_1^+)$ terminates because, as $BinProg$ is a finite set of binary clauses, the loop at lines 2–4 terminates. Concerning $\texttt{satisfy\_DN2–4}$, the inner loops terminate since for each $p \in \Pi$, $Dom(\tau_2^+(p)) \subseteq [1, arity(p)]$ and the outer loop terminates as $BinProg$ is a finite set of binary clauses.

2. - `satisfy_DN1`($BinProg, \tau_1^+$) $\preccurlyeq \tau_1^+$:

     Line 1, we start from $\tau_1^+$. Line 4, we have for each relation symbol $p$ from the heads of the clauses of $BinProg$:

     $$Dom(\tau_2^+(p)) \subseteq Dom(\tau_1^+(p)) \text{ and } \forall i \in Dom(\tau_2^+(p)),\ \tau_2^+(p)(i) = \tau_1^+(p)(i) .$$

     Hence, when we reach line 5, we have: `satisfy_DN1`($BinProg, \tau_1^+$) $\preccurlyeq \tau_1^+$.

   - `satisfy_DN2`($BinProg, \tau_1^+$) $\preccurlyeq \tau_1^+$:

     Line 1, we start from $\tau_1^+$. Then, for each relation symbol $p$ from the heads of the clauses of $BinProg$ and for each $i \in Dom(\tau_2^+(p))$, either $\tau_2^+(p)(i)$ is set to a less general term than $\tau_2^+(p)(i)$ (line 7) or $i$ is removed from the domain of $\tau_2^+(p)$ (lines 6 and 8). Therefore, when we reach line 9, we have: `satisfy_DN2`($BinProg, \tau_1^+$) $\preccurlyeq \tau_1^+$.

   - `satisfy_DN3`($BinProg, \tau_1^+$) $\preccurlyeq \tau_1^+$:

     Line 1, we start from $\tau_1^+$. Line 6, we have for each relation symbol $q$ from the bodies of the clauses of $BinProg$:

     $$Dom(\tau_2^+(q)) \subseteq Dom(\tau_1^+(q)) \text{ and } \forall i \in Dom(\tau_2^+(q)),\ \tau_2^+(q)(i) = \tau_1^+(q)(i) .$$

     Hence, when we reach line 7, we have: `satisfy_DN3`($BinProg, \tau_1^+$) $\preccurlyeq \tau_1^+$.

   - `satisfy_DN4`($BinProg, \tau_1^+$) $\preccurlyeq \tau_1^+$:

     Line 1, we start from $\tau_1^+$. Line 7, we have for each relation symbol $p$ from the heads of the clauses of $BinProg$:

     $$Dom(\tau_2^+(p)) \subseteq Dom(\tau_1^+(p)) \text{ and } \forall i \in Dom(\tau_2^+(p)),\ \tau_2^+(p)(i) = \tau_1^+(p)(i) . \qquad (2)$$

     Hence, when we reach line 8, we have: `satisfy_DN4`($BinProg, \tau_1^+$) $\preccurlyeq \tau_1^+$.

3. Each call to `satisfy_DN1`, …, `satisfy_DN4` terminate. Moreover, concerning function `satisfy_DN4`, we mentioned above that (2) holds. As $\subset$ is a well-founded relation over the set of sets, the loop at lines 5–6 terminates.

4. Line 1, we start from $\tau_1^+$. Then `satisfy_DN1`, …, `satisfy_DN4` weaken $\tau_1^+$ with respect to Definition 3.35. When we reach the fixpoint, the property holds. □

## A.2  Loop Dictionaries

### A.2.1  Proof of Proposition 4.3, page 14

We proceed by induction on the length $n$ of $BinSeq$.

- Basis. If $n = 0$, then, as $([H \leftarrow B], \tau^+)$ is a looping pair, $B$ is $\Delta[\tau^+]$-more general than $H$ and $\tau^+$ is DN for $H \leftarrow B$, *i.e.* $\Delta[\tau^+]$ is DN for $H \leftarrow B$ by Theorem 3.39. Consequently, by Proposition 3.20, $H$ loops w.r.t. $[H \leftarrow B]$.

- Induction. Suppose that for an $n \geq 0$, each looping pair $([H \leftarrow B|BinSeq], \tau^+)$ with $BinSeq$ of length $n$ is such that $H$ loops w.r.t. $[H \leftarrow B|BinSeq]$.

  If $BinSeq$ is of length $n + 1$, it has form $[H_1 \leftarrow B_1|BinSeq_1]$ with $BinSeq_1$ of length $n$. Moreover, as $([H \leftarrow B|BinSeq], \tau^+)$ is a looping pair, there exists a set of positions with associated terms $\tau_1^+$ such that $([H_1 \leftarrow B_1|BinSeq_1], \tau_1^+)$ is a looping pair and $B$ is $\Delta[\tau_1^+]$-more general than $H_1$. So, by the induction hypothesis, $H_1$ loops w.r.t. $[H_1 \leftarrow B_1|BinSeq_1]$ *i.e.* $H_1$ loops w.r.t. $BinSeq$. As $B$ is $\Delta[\tau_1^+]$-more general than $H_1$ and $\Delta[\tau_1^+]$ is DN for $BinSeq$, by Definition 3.18 $B$ loops w.r.t. $BinSeq$. Therefore, by Corollary 3.3, $H$ loops w.r.t. $[H \leftarrow B|BinSeq]$.

### A.2.2 Computing a Loop Dictionary

The top-level function for inferring loop dictionaries from a logic program is the following. It uses the auxiliary functions `unit_loop` and `loops_from_dict` described below.

---

`infer_loop_dict(P, max)`:

**in**: $P$: a logic program
$max$: a non-negative integer

**out**: a loop dictionary, each element $(BinSeq, \tau^+)$ of which
is such that $BinSeq \subseteq T_P^\beta \uparrow max$

1: $Dict := \varnothing$
2: **for each** $H \leftarrow B \in T_P^\beta \uparrow max$ **do**
3: $\quad Dict := \text{unit\_loop}(H \leftarrow B, Dict)$
4: $\quad Dict := \text{loops\_from\_dict}(H \leftarrow B, Dict)$
5: **return** $Dict$

---

The function `unit_loop` is used to extract from a binary clause the most simple form of a looping pair:

---

`unit_loop(H ← B, Dict)`:

**in**: $H \leftarrow B$: a binary clause
$Dict$: a loop dictionary

**out**: $Dict'$: a loop dictionary, every element $(BinSeq, \tau^+)$ of which is
such that either $(BinSeq, \tau^+) \in Dict$ or $BinSeq = [H \leftarrow B]$

1: $Dict' := Dict$
2: $\tau^+ := \text{dna}([H \leftarrow B], \tau_{max}^+)$
3: **if** $B$ is $\Delta[\tau^+]$-more general than $H$ **then**
4: $\quad Dict' := Dict' \cup \{([H \leftarrow B], \tau^+)\}$
5: **return** $Dict'$

---

Termination of `unit_loop` relies on that of `dna`. Partial correctness is deduced from the next theorem.

**Theorem A.3** *(Partial correctness of* `unit_loop`*) Let $H \leftarrow B$ be a binary clause and $Dict$ be a loop dictionary. Then* `unit_loop`*$(H \leftarrow B, Dict)$ is a loop dictionary, every element $(BinSeq, \tau^+)$ of which is such that either $(BinSeq, \tau^+) \in Dict$ or $BinSeq = [H \leftarrow B]$.*

*Proof.* Let $\tau^+$ be the set of positions with associated terms computed at line 2. If $B$ is not $\Delta[\tau^+]$-more general than $H$ then, at line 5 of `unit_loop`, we have $Dict' = Dict$, so the theorem holds.

Now suppose that $B$ is $\Delta[\tau^+]$-more general than $H$. Then, at line 5 we have $Dict' := Dict \cup \{([H \leftarrow B], \tau^+)\}$ where $Dict$ is a loop dictionary, $\tau^+$ is DN for $H \leftarrow B$ and $B$ is $\Delta[\tau^+]$-more general than $H$. So at line 5 $Dict'$ is a loop dictionary, every element $(BinSeq, \tau^+)$ of which is such that either $(BinSeq, \tau^+) \in Dict$ or $BinSeq = [H \leftarrow B]$.

The function `loops_from_dict` is used to combine a binary clause to some looping pairs that have already been infered in order to get some more looping pairs.

```
loops_from_dict(H ← B, Dict):
   in:  H ← B: a binary clause
        Dict: a loop dictionary
   out: Dict′: a loop dictionary, every element (BinSeq, τ⁺) of which is
        such that (BinSeq, τ⁺) ∈ Dict or BinSeq = [H ← B|BinSeq₁]
        for some (BinSeq₁, τ₁⁺) in Dict
   1:  Dict′ := Dict
   2:  for each ([H₁ ← B₁|BinSeq₁], τ₁⁺) ∈ Dict do
   3:     if B is Δ[τ₁⁺]-more general than H₁ then
   4:        τ⁺ := dna([H ← B, H₁ ← B₁|BinSeq₁], τ₁⁺)
   5:        Dict′ := Dict′ ∪ {([H ← B, H₁ ← B₁|BinSeq₁], τ⁺)}
   6:  return Dict′
```

Termination of $\texttt{loops\_from\_dict}$ follows from finiteness of $Dict$ (because $Dict$ is a loop dictionary) and termination of $\texttt{dna}$. Partial correctness follows from the result below.

**Theorem A.4** (*Partial correctness of* $\texttt{loops\_from\_dict}$) *Let* $H \leftarrow B$ *be a binary clause and* $Dict$ *be a loop dictionary. Then* $\texttt{loops\_from\_dict}(H \leftarrow B, Dict)$ *is a loop dictionary, every element* $(BinSeq, \tau^+)$ *of which is such that* $(BinSeq, \tau^+) \in Dict$ *or* $BinSeq = [H \leftarrow B|BinSeq_1]$ *for some* $(BinSeq_1, \tau_1^+)$ *in* $Dict$.

*Proof.* Upon initialization at line 1, $Dict'$ is a loop dictionary. Suppose that before an iteration of the loop at line 2, $Dict'$ is a loop dictionary. Let $([H_1 \leftarrow B_1|BinSeq_1], \tau_1^+) \in Dict$.

If the condition at line 3 is false, then $Dict'$ remains unchanged, so after the iteration $Dict'$ is still a loop dictionary. Otherwise, the pair $([H \leftarrow B, H_1 \leftarrow B_1|BinSeq_1], \tau^+)$ is added to $Dict'$ at line 5. Notice that this pair is a looping one because $\tau^+$ defined at line 4 is DN for $[H \leftarrow B, H_1 \leftarrow B_1|BinSeq_1]$ and $([H_1 \leftarrow B_1|BinSeq_1], \tau_1^+)$ is a looping pair and $B$ is $\Delta[\tau_1^+]$-more general than $H_1$. Therefore, after the iteration, $Dict'$ is a loop dictionary. Finally, notice that as $Dict$ is a finite set, the loop at line 2 terminates. Hence, at line 6 $Dict'$ is a finite set of looping pairs *i.e.* $Dict'$ is a loop dictionary.

Moreover, at line 1, each element of $Dict'$ belongs to $Dict$. Then, during the loop, pairs of form $([H \leftarrow B|BinSeq_1], \tau^+)$ are added to $Dict'$ where $BinSeq_1$ is such that there exists $(BinSeq_1, \tau_1^+) \in Dict$. Consequently, at line 6 each element $(BinSeq, \tau^+)$ of $Dict'$ is such that either $(BinSeq, \tau^+) \in Dict$ or $BinSeq = [H \leftarrow B|BinSeq_1]$ for some $(BinSeq_1, \tau_1^+)$ in $Dict$.

Finally, here is the correctness proof of the function $\texttt{infer\_loop\_dict}$.

**Theorem A.5** (*Correctness of* $\texttt{infer\_loop\_dict}$) *Let* $P$ *be a logic program and* $max$ *be a non-negative integer. Then,* $\texttt{infer\_loop\_dict}(P, max)$ *terminates and returns a loop dictionary, every element* $(BinSeq, \tau^+)$ *of which is such that* $BinSeq \subseteq T_P^\beta \uparrow max$.

*Proof.* At line 1, $Dict$ is initialized to $\varnothing$ which is a loop dictionary. Suppose that before an iteration of the loop at line 2, $Dict$ is a loop dictionary. Then at lines 3 and 4 $\texttt{unit\_loop}$ and $\texttt{loops\_from\_dict}$ fullfil their specifications. Hence, the call to these functions terminates and after the iteration $Dict$ is still a loop dictionary. Finally, as $T_P^\beta \uparrow max$ is a finite set, the loop at line 2 terminates and at line 5 $Dict$ is a loop dictionary.

Moreover, at line 1 each element $(BinSeq, \tau^+)$ of $Dict$ is such that $BinSeq \subseteq T_P^\beta \uparrow max$. Then, during the loop, $\texttt{unit\_loop}$ and $\texttt{loops\_from\_dict}$ are called with clauses from $T_P^\beta \uparrow max$. So, by Theorem A.3 and Theorem A.4, after the iteration each element $(BinSeq, \tau^+)$ of $Dict$ is such that $BinSeq \subseteq T_P^\beta \uparrow max$.

## A.3  Looping Conditions

The following function computes a finite set of looping conditions for any given logic program.

---

```
infer_loop_cond(P, max):

   in:  P: a logic program
        max: a non-negative integer

  out:  a finite set of looping conditions for P

   1:  L := ∅
   2:  Dict := infer_loop_dict(P, max)
   3:  for each ([H ← B|BinSeq], τ⁺) ∈ Dict do
   4:      L := L ∪ {(H, τ⁺)}
   5:  return L
```

---

A call to `infer_loop_cond`$(P, max)$ terminates for any logic program $P$ and any non-negative integer $max$ because `infer_loop_dict`$(P, max)$ at line 2 terminates and the loop at line 3 has a finite number of iterations (because, by correctness of `infer_loop_dict`, $Dict$ is finite.) Partial correctness of `infer_loop_cond` follows from the next theorem.

**Theorem A.6** (*Partial correctness of* `infer_loop_cond`) *Let $P$ be a logic program and $max$ be a non-negative integer. Then* `infer_loop_cond`$(P, max)$ *is a finite set of looping conditions for $P$.*

*Proof.* By correctness of `infer_loop_dict`, $Dict$ is a loop dictionary.

Let $([H \leftarrow B|BinSeq], \tau^+) \in Dict$. Then $([H \leftarrow B|BinSeq], \tau^+)$ is a looping pair. Consequently, by Proposition 4.3, $H$ loops w.r.t. $[H \leftarrow B|BinSeq]$. As $\tau^+$, hence $\Delta[\tau^+]$, is DN for $[H \leftarrow B|BinSeq]$, by Definition 3.18 every atom that is $\Delta[\tau^+]$-more general than $H$ loops w.r.t. $[H \leftarrow B|BinSeq]$.

As $([H \leftarrow B|BinSeq], \tau^+) \in Dict$, by Theorem A.5 we have

$$[H \leftarrow B|BinSeq] \subseteq T_P^\beta \uparrow max \subseteq bin\_unf(P) \ .$$

So, by Theorem 2.1, $H$ left loops w.r.t. $P$ and every atom that is $\Delta[\tau^+]$-more general than $H$ left loops w.r.t. $P$. So, $(H, \tau^+)$ is a looping condition for $P$. Consequently, at line 5, $L$ is a finite set of looping conditions for $P$ because, as $Dict$ is finite, the loop at line 3 iterates a finite number of times.

# B   Proofs

## B.1   Two Useful Lemmas

**Lemma B.1** *Let $c := H \leftarrow B$ be a binary clause. Then, for every variant $c\gamma$ of $c$ such that $Var(c\gamma) \cap Var(H) = \varnothing$, we have $H \underset{c}{\Longrightarrow} B\gamma'$ where $\gamma' := \gamma | Var(B) \setminus Var(H)$.*

*Proof.* Let $\mu := \{x\gamma/x | x \in Var(H)\}$. By Claim B.2 below, $\mu$ is an mgu of $H\gamma$ and $H$. Hence, as $Var(c\gamma) \cap Var(H) = \varnothing$, we have the left derivation step $H \overset{\mu}{\underset{c}{\Longrightarrow}} B\gamma\mu$ where $c\gamma$ is the input clause used.

If $Var(B) = \varnothing$, then $B\gamma\mu = B\gamma'$, so we have $H \overset{\mu}{\underset{c}{\Longrightarrow}} B\gamma'$ *i.e.* $H \underset{c}{\Longrightarrow} B\gamma'$.

If $Var(B) \neq \varnothing$, take a variable $x \in Var(B)$:

- if $x \in Var(H)$, then $x(\gamma\mu) = (x\gamma)\mu = x$ by definition of $\mu$,

- if $x \notin Var(H)$, then $x(\gamma\mu) = (x\gamma)\mu = x\gamma$ by definition of $\mu$.

Hence, $B\gamma\mu = B\gamma'$, so we have $H \overset{\mu}{\underset{c}{\Longrightarrow}} B\gamma'$ *i.e.* $H \underset{c}{\Longrightarrow} B\gamma'$.

**Claim B.2** $\mu$ *is an mgu of $H\gamma$ and $H$.*

*Proof.* Let $p(s_1, \ldots, s_n) := H$. The set of unifiers of $H\gamma$ and $H$ is the same as that of $E_1 := \{s_1\gamma = s_1, \ldots, s_n\gamma = s_n\}$. Let $E_2 := \{x\gamma = x \mid x \in Var(H)\}$. Notice that, as $\gamma$ is a renaming, if $x, y \in Var(H)$ then $x \neq y \Rightarrow x\gamma \neq y\gamma$. Moreover, for each $x \in Var(H)$, $x\gamma \neq x$ because $Var(c\gamma) \cap Var(H) = \varnothing$. So, $E_2$ is solved. Consequently, by Lemma 2.15 page 32 of [1], $\mu$ is an mgu of $E_2$. Notice that, by Claim B.3 below, the set of unifiers of $E_1$ is that of $E_2$. So $\mu$ is an mgu of $E_1$ *i.e.* $\mu$ is an mgu of $H\gamma$ and $H$.

**Claim B.3** *$E_1$ and $E_2$ have the same set of unifiers.*

*Proof.* Let $\theta$ be a unifier of $E_1$. Let $x \in Var(H)$ and let $i \in [1, n]$ such that $x \in Var(s_i)$. Then $s_i\gamma\theta = s_i\theta$, so, if $x_k$ is an occurrence of $x$ in $s_i$, we have $x_k\gamma\theta = x_k\theta$ *i.e.* $(x_k\gamma)\theta = x_k\theta$. As $x$ denotes any variable of $H$, we conclude that $\theta$ is a unifier of $E_2$. Conversely, let $\theta$ be a unifier of $E_2$. Then, for each $i \in [1, n]$, $(s_i\gamma)\theta = s_i\theta$ by definition of $E_2$. Hence, $\theta$ is a unifier of $E_1$.

**Lemma B.4** *Let $c := H \leftarrow B$ be a binary clause, $c\gamma$ be a variant of $c$ such that $Var(c\gamma) \cap Var(H) = \varnothing$ and $\gamma' := \gamma | Var(B) \setminus Var(H)$. Then, there exists a renaming $\gamma''$ such that $B\gamma' = B\gamma''$.*

*Proof.* Let $\mathcal{A} := \{x \mid x \in Ran(\gamma')$ and $x \notin Dom(\gamma')\}$ and $\mathcal{B} := \{x \mid x \in Dom(\gamma')$ and $x \notin Ran(\gamma')\}$. Notice that $Ran(\gamma')$ and $Dom(\gamma')$ have the same number of elements, so $\mathcal{A}$ and $\mathcal{B}$ have the same number of elements. Let $\sigma$ be a 1-1 and onto mapping from $\mathcal{A}$ to $\mathcal{B}$. Then, $\gamma'' := \gamma' \cup \sigma$ is a well-defined substitution, is such that $Dom(\gamma'') = Ran(\gamma'')$, is 1-1 and is onto. Consequently, $\gamma''$ is a renaming.

Now, let us prove that $B\gamma' = B\gamma''$. If $Var(B) = \varnothing$, then the result is straightforward. Otherwise, let $x \in Var(B)$.

- If $x \in Var(H)$ then, as $Dom(\gamma') \subseteq Var(B) \setminus Var(H)$, we have $x \notin Dom(\gamma')$, so $x\gamma' = x$. Moreover, $x\gamma'' = x(\gamma' \cup \sigma) = x\sigma = x$ because $Dom(\sigma) \subseteq Ran(\gamma')$ and $Ran(\gamma') \cap Var(H) = \varnothing$. Consequently, we have $x\gamma' = x\gamma''$.

- If $x \notin Var(H)$ and $x \in Dom(\gamma')$ then $x\gamma'' = x(\gamma' \cup \sigma) = x\gamma'$.

- If $x \notin Var(H)$ and $x \notin Dom(\gamma')$ then $x\gamma' = x$.

  Now, suppose that $x \in Dom(\sigma)$. Then, as $Dom(\sigma) \subseteq Ran(\gamma') \subseteq Ran(\gamma)$, we have $x \in Ran(\gamma)$. But, as $\gamma$ is a renaming, $Ran(\gamma) = Dom(\gamma)$, so we have $x \in Dom(\gamma)$. As $x \in Var(B)$, as $x \notin Var(H)$ and as $\gamma' := \gamma | Var(B) \setminus Var(H)$, we have $x \in Dom(\gamma')$. Contradiction.

  Consequently, $x \notin Dom(\sigma)$, so $x\sigma = x$ and $x\gamma'' = x(\gamma' \cup \sigma) = x\sigma = x$. Finally, we have proved that $x\gamma' = x\gamma''$.

## B.2 Proof of Corollary 3.2, page 4

By Lemma B.1 and Lemma B.4, we have $H \underset{c}{\Longrightarrow} B\gamma''$ where $\gamma''$ is a renaming. As by hypothesis $B$ is more general than $H$, then $B\gamma''$ is more general than $H$. Therefore, by the One Step Lifting Lemma 3.1, $H$ loops w.r.t. $\{c\}$.

## B.3 Proof of Corollary 3.3, page 4

By Lemma B.1 and Lemma B.4, we have $H \underset{c}{\Longrightarrow} B\gamma''$ where $\gamma''$ is a renaming. As $B\gamma''$ is more general than $B$ and as $B$ loops w.r.t. $P$, then, by the One Step Lifting Lemma 3.1, $B\gamma''$ loops w.r.t. $P$, so $H$ loops w.r.t. $P$.

### B.4  Proof of Proposition 3.16, page 7

If $A$ is $\Delta$-more general than $B$, we have, for a substitution $\eta$:

$$\left\{ \begin{array}{l} A = p(s_1, \ldots, s_n) \\ B = p(t_1, \ldots, t_n) \\ \forall i \in [1, n] \setminus Dom(\Delta(p)), \ t_i = s_i \eta \\ \forall i \in Dom(\Delta(p)), \ \Delta(p)(i)(s_i) = \mathtt{true} . \end{array} \right.$$

Let $A'$ be a variant of $A$. Then, there exists a renaming $\gamma$ such that $A' = A\gamma$. As for each $i \in Dom(\Delta(p))$, $\Delta(p)(i)$ is a variant independent term-condition, we have:

$$\left\{ \begin{array}{l} A' = p(s_1\gamma, \ldots, s_n\gamma) \\ B = p(t_1, \ldots, t_n) \\ \forall i \in [1, n] \setminus Dom(\Delta(p)), \ t_i = s_i \eta = (s_i\gamma)(\gamma^{-1}\eta) \\ \forall i \in Dom(\Delta(p)), \ \Delta(p)(i)(s_i\gamma) = \mathtt{true} . \end{array} \right.$$

Consequently, $A'$ is $\Delta$-more general than $B$ for $\gamma^{-1}\eta$, *i.e.* $A'$ is $\Delta$-more general than $B$.

### B.5  Proof of Proposition 3.17, page 7

$\Leftarrow$ By definition.

$\Rightarrow$ Let $p(s_1, \ldots, s_n) := A$ and $p(t_1, \ldots, t_n) := B$. As $A$ is $\Delta$-more general than $B$, there exists a substitution $\sigma$ such that $A$ is $\Delta$-more general than $B$ for $\sigma$. Notice that $A$ is also $\Delta$-more general than $B$ for the substitution obtained by restricting the domain of $\sigma$ to the variables appearing in the positions of $A$ not distinguished by $\Delta$. More precisely, let

$$\eta := \sigma|Var(\{s_i \mid i \in [1, n] \setminus Dom(\Delta(p))\}) .$$

Then,

$$Dom(\eta) \subseteq Var(A) \tag{3}$$

and $A$ is $\Delta$-more general than $B$ for $\eta$.

Now, let $x \in Dom(\eta)$. Then, there exists $i \in [1, n] \setminus Dom(\Delta(p))$ such that $x \in Var(s_i)$. As $A$ is $\Delta$-more general than $B$ for $\eta$ and $i \in [1, n] \setminus Dom(\Delta(p))$, we have $t_i = s_i\eta$. So, as $x \in Var(s_i)$, $x\eta$ is a subterm of $t_i$. Consequently, $Var(x\eta) \subseteq Var(t_i)$, so $Var(x\eta) \subseteq Var(B)$.

So, we have proved that for each $x \in Dom(\eta)$, $Var(x\eta) \subseteq Var(B)$, *i.e.* we have proved that

$$Ran(\eta) \subseteq Var(B) . \tag{4}$$

Finally, (3) and (4) imply that $Dom(\eta) \cup Ran(\eta) \subseteq Var(A, B)$ *i.e.* that

$$Var(\eta) \subseteq Var(A, B) .$$

### B.6  Proof of Proposition 3.20, page 8

By Lemma B.1 and Lemma B.4, we have $H \underset{c}{\Longrightarrow} B\gamma''$ where $\gamma''$ is a renaming. As by hypothesis $B$ is $\Delta$-more general than $H$, then by Proposition 3.16 $B\gamma''$ is $\Delta$-more general than $H$. Therefore, as $\Delta$ is DN for $c$, by Definition 3.18, $H$ loops w.r.t. $\{c\}$.

### B.7  Proof of Proposition 3.37, page 11

Let $c := p(s_1, \ldots, s_n) \leftarrow q(t_1, \ldots, t_m)$ and $c' := p(s'_1, \ldots, s'_n) \leftarrow q(t'_1, \ldots, t'_m)$ be a variant of $c$. Then, there exists a renaming $\gamma$ such that $c' = c\gamma$.

**(DN1)** Let $i \in Dom(\tau^+(p))$. Suppose that there exists $j \neq i$ such that $Var(s'_i) \cap Var(s'_j) \neq \varnothing$ and let us derive a contradiction.

Let $x' \in Var(s'_i) \cap Var(s'_j)$. As $s'_j = s_j \gamma$, there exists $x \in Var(s_j)$ such that $x' = x\gamma$.

For such an $x$, as $j \neq i$ and as $Var(s_i) \cap Var(s_j) = \varnothing$ (because $\tau^+$ is DN for $c$), we have $x \notin Var(s_i)$. So, as $\gamma$ is a 1-1 and onto mapping from its domain to itself, we have $x\gamma \notin Var(s_i\gamma)$[3], *i.e.* $x' \notin Var(s'_i)$. Contradiction!

Consequently, $Var(s'_i) \cap Var(s'_j) = \varnothing$.

**(DN2)** Let $\langle i \mapsto u_i \rangle \in \tau^+(p)$. As $s_i$ is more general than $u_i$ (because $\tau^+$ is DN for $c$) and as $s'_i$ is a variant of $s_i$, $s'_i$ is more general than $u_i$.

**(DN3)** Let $\langle j \mapsto u_j \rangle \in \tau^+(q)$. As $t_j$ is an instance of $u_j$ (because $\tau^+$ is DN for $c$) and as $t'_j$ is a variant of $t_j$, $t'_j$ is an instance of $u_j$.

**(DN4)** Let $i \in Dom(\tau^+(p))$. Suppose there exists $j \notin Dom(\tau^+(q))$ such that $Var(s'_i) \cap Var(t'_j) \neq \varnothing$. Let us derive a contradiction.

Let $x' \in Var(s'_i) \cap Var(t'_j)$. As $t'_j = t_j\gamma$ and $x' \in Var(t'_j)$, there exists $x \in Var(t_j)$ such that $x' = x\gamma$. For such an $x$, as the elements of $Var(s_i)$ only occur in those $t_k$ s.t. $k \in Dom(\tau^+(q))$ (because $\tau^+$ is DN for $c$) and as $x \in Var(t_j)$ with $j \notin Dom(\tau^+(q))$, we have $x \notin Var(s_i)$. So, as $\gamma$ is a 1-1 and onto mapping from its domain to itself, we have $x\gamma \notin Var(s_i\gamma)$ (see footnote 3), *i.e.* $x' \notin Var(s'_i)$. Contradiction! So, for each $j \notin Dom(\tau^+(q))$, we have $Var(s'_i) \cap Var(t'_j) = \varnothing$.

Finally, we have established that $\tau^+$ is DN for $c'$.

## B.8 DN Sets of Positions with Associated Terms Generate DN Filters

In this section, we give a proof of Theorem 3.39, page 12.

### B.8.1 Context

All the results of this section are parametric to the following context:

- $P$ is a binary program and $\tau^+$ is a set of positions with associated terms that is DN for $P$,

- $Q \stackrel{\theta}{\underset{c}{\Longrightarrow}} Q_1$ is a left derivation step where

  - $c \in P$,
  - $Q := p(t_1, \ldots, t_n)$,
  - $c_1 := p(s_1, \ldots, s_n) \leftarrow B$ is the input clause used (consequently, $c_1$ is a variant of $c$ that is variable disjoint from $Q$),

- $Q' := p(t'_1, \ldots, t'_n)$ is $\Delta[\tau^+]$-more general than $Q$ *i.e.*, by Proposition 3.17, there exists a substitution $\eta$ such that $Var(\eta) \subseteq Var(Q, Q')$ and $Q'$ is $\Delta[\tau^+]$-more general than $Q$ for $\eta$.

---

[3] Because if $x\gamma \in Var(s_i\gamma)$, then either $x \in Var(s_i)$, either $x\gamma \in Var(s_i)$ and $(x\gamma)\gamma = x\gamma$. The former case is impossible because we said that $x \notin Var(s_i)$. The latter case is impossible too because $(x\gamma)\gamma = x\gamma$ implies that $x\gamma \notin Dom(\gamma)$ *i.e.* $x \notin Dom(\gamma)$ (because $\gamma$ is a 1-1 and onto mapping from its domain to itself); so, $x = x\gamma$ *i.e.*, as $x\gamma \in Var(s_i)$, $x \in Var(s_i)$.

### B.8.2 Technical Definitions and Lemmas

**Definition B.5** *(Technical Definition) Let $c_1' := p(s_1', \ldots, s_n') \leftarrow B'$ be a binary clause such that*

- $Var(c_1') \cap Var(Q, Q') = \varnothing$ *and*

- $c_1 = c_1'\gamma$ *for some renaming $\gamma$ satisfying $Var(\gamma) \subseteq Var(c_1, c_1')$.*

*As $c_1'$ is a variant of $c_1$ and $c_1$ is a variant of $c$, then $c_1'$ is a variant of $c$. Moreover, as $\tau^+$ is DN for $c$, by Proposition 3.37, $\tau^+$ is DN for $c_1'$. So, by **(DN2)** in Definition 3.35, for each $\langle i \mapsto u_i \rangle \in \tau^+(p)$ there exists a substitution $\delta_i$ such that $u_i = s_i'\delta_i$.*

*Moreover, as $p(t_1', \ldots, t_n')$ is $\Delta[\tau^+]$-more general than $p(t_1, \ldots, t_n)$, for each $\langle i \mapsto u_i \rangle \in \tau^+(p)$, $t_i'$ is an instance of $u_i$. So, there exists a substitution $\delta_i'$ such that $t_i' = u_i\delta_i'$.*

*For each $i \in Dom(\tau^+(p))$, we set*

$$\sigma_i \stackrel{def}{=} (\delta_i\delta_i')|Var(s_i') .$$

*Moreover, we set:*

$$\sigma \stackrel{def}{=} \bigcup_{i \in Dom(\tau^+(p))} \sigma_i .$$

**Lemma B.6** *The set $\sigma$ of Definition B.5 is a well-defined substitution.*

*Proof.* Notice that, as $\tau^+$ is DN for $c_1'$, by **(DN1)** in Definition 3.35, we have

$$\forall i \in Dom(\tau^+(p)), \ \forall j \in [1, n] \setminus \{i\}, \ Var(s_i') \cap Var(s_j') = \varnothing .$$

Consequently,

$$\forall i, j \in Dom(\tau^+(p)), \ i \neq j \Rightarrow Dom(\sigma_i) \cap Dom(\sigma_j) = \varnothing .$$

Moreover, for each $i \in Dom(\tau^+(p))$, $\sigma_i$ is a well-defined substitution. So, $\sigma$ is a well-defined substitution. $\qquad\blacksquare$

**Lemma B.7** *(Technical Lemma) Let $c_1' := p(s_1', \ldots, s_n') \leftarrow B'$ be a binary clause such that*

- $Var(c_1') \cap Var(Q, Q') = \varnothing$ *and*

- $c_1 = c_1'\gamma$ *for some renaming $\gamma$ satisfying $Var(\gamma) \subseteq Var(c_1, c_1')$.*

*Let $\sigma$ be the substitution of Definition B.5. Then, the substitution $\sigma\eta\gamma\theta$ is a unifier of $p(t_1', \ldots, t_n')$ and $p(s_1', \ldots, s_n')$.*

*Proof.* The result follows from the following facts.

- For each $\langle i \mapsto u_i \rangle \in \tau^+(p)$, we have:

$$s_i'\sigma = s_i'\sigma_i = s_i'\delta_i\delta_i' = (s_i'\delta_i)\delta_i' = u_i\delta_i' = t_i'$$

  and $t_i'\sigma = t_i'$ because $Dom(\sigma) \subseteq Var(c_1')$ and $Var(Q') \cap Var(c_1') = \varnothing$. So, $s_i'\sigma = t_i'\sigma$ and $s_i'\sigma\eta\gamma\theta = t_i'\sigma\eta\gamma\theta$.

- For each $i \in [1, n] \setminus Dom(\tau^+(p))$, we have:

$$s_i'\eta\gamma\theta = (s_i'\eta)\gamma\theta = s_i'\gamma\theta = (s_i'\gamma)\theta = s_i\theta$$

  and

$$t_i'\eta\gamma\theta = (t_i'\eta)\gamma\theta = t_i\gamma\theta = (t_i\gamma)\theta = t_i\theta$$

  and $s_i\theta = t_i\theta$ because $\theta$ is a unifier of $p(s_1, \ldots, s_n)$ and $p(t_1, \ldots, t_n)$ (because $Q \stackrel{\theta}{\underset{c}{\Longrightarrow}} Q_1$ with $c_1$ as input clause used). So,

$$s_i'\eta\gamma\theta = t_i'\eta\gamma\theta \qquad (5)$$

- For each $i \in [1,n] \setminus Dom(\tau^+(p))$, we also have:

  – $s_i'\sigma = s_i'$ because $Dom(\sigma) = Var(\{s_j' \mid j \in Dom(\tau^+(p))\})$ and, by **(DN1)** in Definition 3.35, $Var(\{s_j' \mid j \in Dom(\tau^+(p))\}) \cap Var(s_i') = \varnothing$;

  – $t_i'\sigma = t_i'$ because $Dom(\sigma) \subseteq Var(c_1')$ and $Var(Q') \cap Var(c_1') = \varnothing$.

  Therefore, because of (5), $s_i'\sigma\eta\gamma\theta = t_i'\sigma\eta\gamma\theta$.

### B.8.3  $\Delta$-Propagation

Now we extend, in the case of left derivations with atomic queries and binary clauses, the following Propagation Lemma that is proved by Apt in [1] p. 54–56.

**Lemma B.8** *(Propagation) Let $G$, $G_1$, $G'$ and $G_1'$ be some queries such that*

$$G \underset{c}{\Longrightarrow} G_1 \quad \text{and} \quad G' \underset{c}{\Longrightarrow} G_1' \quad \text{and}$$

- *$G$ is an instance of $G'$*

- *in $G$ and $G'$ atoms in the same positions are selected.*

*Then, $G_1$ is an instance of $G_1'$.*

First we establish the following result.

**Lemma B.9** *Suppose there exists a left derivation step of form $Q' \underset{c}{\overset{\theta'}{\Longrightarrow}} Q_1'$ where the input clause is $c_1'$ such that $Var(Q) \cap Var(c_1') = \varnothing$. Then, $Q_1'$ is $\Delta[\tau^+]$-more general than $Q_1$.*

*Proof.* Notice that we have

$$Var(Q) \cap Var(c_1) = Var(Q, Q') \cap Var(c_1') = \varnothing .$$

Moreover, as $c_1$ is a variant of $c_1'$, there exists a renaming $\gamma$ such that

$$Var(\gamma) \subseteq Var(c_1, c_1') \quad \text{and} \quad c_1 = c_1'\gamma .$$

Let $c_1' := p(s_1', \ldots, s_n') \leftarrow B'$. Then,

$$Q_1 = B\theta \quad \text{and} \quad Q_1' = B'\theta' .$$

$\tau^+$ is DN for $c$ and $c_1'$ is a variant of $c$. So, by Proposition 3.37, $\tau^+$ is DN for $c_1'$. Let $\sigma$ be the substitution of Definition B.5.

Let $q(v_1', \ldots, v_m') := B'$. As $B = B'\gamma$, $B$ has form $q(v_1, \ldots, v_m)$.

- For each $\langle j \mapsto u_j \rangle \in \tau^+(q)$, $v_j'$ is an instance of $u_j$ (because $\tau^+$ is DN for $c_1'$ and **(DN3)** in Definition 3.35.)

- For each $j \in [1,m] \setminus Dom(\tau^+(q))$ we have:

$$v_j'\sigma\eta\gamma\theta = (v_j'\sigma)\eta\gamma\theta = v_j'\eta\gamma\theta$$

because, by **(DN4)** in Definition 3.35

$$Var(v_j') \cap Var(\{s_i' \mid i \in Dom(\tau^+(p))\}) = \varnothing$$

with $Dom(\sigma) = Var(\{s_i' \mid i \in Dom(\tau^+(p))\})$. Moreover,

$$v_j'\eta\gamma\theta = (v_j'\eta)\gamma\theta = v_j'\gamma\theta$$

because $Var(\eta) \subseteq Var(Q, Q')$ and $Var(c_1') \cap Var(Q, Q') = \varnothing$. Finally,

$$v_j'\gamma\theta = (v_j'\gamma)\theta = v_j\theta$$

because $B = B'\gamma$.

Consequently, we have proved that

$$q(v'_1, \ldots, v'_m) \text{ is } \Delta[\tau^+]\text{-more general than } q(v_1, \ldots, v_m)\theta \text{ for } \sigma\eta\gamma\theta$$

*i.e.* that $B'$ is $\Delta[\tau^+]$-more general than $B\theta$ for $\sigma\eta\gamma\theta$ *i.e.* that

$$B' \text{ is } \Delta[\tau^+]\text{-more general than } Q_1 \text{ for } \sigma\eta\gamma\theta \ . \tag{6}$$

But, by the Technical Lemma B.7, $\sigma\eta\gamma\theta$ is a unifier of $p(s'_1, \ldots, s'_n)$ and $p(t'_1, \ldots, t'_n)$. As $\theta'$ is an mgu of $p(s'_1, \ldots, s'_n)$ and $p(t'_1, \ldots, t'_n)$ (because $Q' \stackrel{\theta'}{\underset{c}{\Longrightarrow}} Q'_1$ with $c'_1$ as input clause), there exists $\delta$ such that $\sigma\eta\gamma\theta = \theta'\delta$. Therefore, we conclude from (6) that $B'$ is $\Delta[\tau^+]$-more general than $Q_1$ for $\theta'\delta$ which implies that $B'\theta'$ is $\Delta[\tau^+]$-more general than $Q_1$ for $\delta$ *i.e.* that $Q'_1$ is $\Delta[\tau^+]$-more general than $Q_1$ for $\delta$. Finally, we have proved that $Q'_1$ is $\Delta[\tau^+]$-more general than $Q_1$.

Using the Propagation Lemma B.8, the preceding result can be extended as follows.

**Proposition B.10** *($\Delta$-Propagation) Suppose there exists a left derivation step $Q' \stackrel{\theta'}{\underset{c}{\Longrightarrow}} Q'_1$. Then $Q'_1$ is $\Delta[\tau^+]$-more general than $Q_1$.*

*Proof.* Let $c'_1$ be the input clause used in $Q' \stackrel{\theta'}{\underset{c}{\Longrightarrow}} Q'_1$. Take a variant $Q''$ of $Q$ such that

$$Var(Q'') \cap Var(c'_1) = \varnothing$$

and a variant $c''_1$ of $c$ such that

$$Var(c''_1) \cap Var(Q'') = \varnothing \ .$$

Then, the left resolvent $Q''_1$ of $Q''$ and $c$ exists with the input clause $c''_1$. So, for some $\theta''$, we have $Q'' \stackrel{\theta''}{\underset{c}{\Longrightarrow}} Q''_1$ with input clause $c''_1$. Consequently, we have:

$$Q \stackrel{\theta}{\underset{c}{\Longrightarrow}} Q_1 \quad \text{and} \quad Q'' \stackrel{\theta''}{\underset{c}{\Longrightarrow}} Q''_1 \ .$$

$Q$ and $Q''$ are instances of each other because $Q''$ is a variant of $Q$. So, by the Propagation Lemma B.8 used twice, $Q''_1$ is an instance of $Q_1$ and $Q_1$ is an instance of $Q''_1$. So,

$$Q''_1 \text{ is a variant of } Q_1 \ . \tag{7}$$

But we also have

$$Q'' \stackrel{\theta''}{\underset{c}{\Longrightarrow}} Q''_1 \quad \text{and} \quad Q' \stackrel{\theta'}{\underset{c}{\Longrightarrow}} Q'_1$$

with input clauses $c''_1$ and $c'_1$, with $Q'$ that is $\Delta[\tau^+]$-more general than $Q''$ (because $Q''$ is a variant of $Q$ and $Q'$ is $\Delta[\tau^+]$-more general than $Q$) and $Var(Q'') \cap Var(c'_1) = \varnothing$. So, by Lemma B.9,

$$Q'_1 \text{ is } \Delta[\tau^+]\text{-more general than } Q''_1 \ . \tag{8}$$

Finally, from (7) and (8) we have: $Q'_1$ is $\Delta[\tau^+]$-more general than $Q_1$.

### B.8.4 Epilogue

Theorem 3.39 is a direct consequence of the following result.

**Proposition B.11** *(One Step $\Delta$-Lifting) Let $c'$ be a variant of $c$ variable disjoint with $Q'$. Then, there exist $\theta'$ and a query $Q'_1$ that is $\Delta[\tau^+]$-more general than $Q_1$ such that $Q' \stackrel{\theta'}{\underset{c}{\Longrightarrow}} Q'_1$ with input clause $c'$.*

*Proof.* Let $c'_1 := p(s'_1, \ldots, s'_n) \leftarrow B'$ be a variant of $c_1$. Then there exists a renaming $\gamma$ such that $Var(\gamma) \subseteq Var(c_1, c'_1)$ and $c_1 = c'_1\gamma$. Suppose also that

$$Var(c'_1) \cap Var(Q, Q') = \varnothing \ .$$

By the Technical Lemma B.7, $p(s'_1, \ldots, s'_n)$ and $p(t'_1, \ldots, t'_n)$ unify. Moreover, as $Var(c'_1) \cap Var(Q') = \varnothing$, $p(s'_1, \ldots, s'_n)$ and $p(t'_1, \ldots, t'_n)$ are variable disjoint. Notice that the following claim holds.

**Claim B.12** *Suppose that the atoms $A$ and $H$ are variable disjoint and unify. Then, $A$ also unifies with any variant $H'$ of $H$ variable disjoint with $A$.*

*Proof.* For some $\gamma$ such that $Dom(\gamma) \subseteq Var(H')$, we have $H = H'\gamma$. Let $\theta$ be a unifier of $A$ and $H$. Then, $A\gamma\theta = A\theta = H\theta = H'\gamma\theta$, so $A$ and $H'$ unify.

Therefore, as $c'$ is a variant of $c'_1$ and $c'$ is variable disjoint with $Q'$, $p(t'_1, \ldots, t'_n)$ and the head of $c'$ unify. As they also are variable disjoint, we have

$$Q' \overset{\theta'}{\underset{c'}{\Longrightarrow}} Q'_1$$

for some $\theta'$ and $Q'_1$ where $c'$ is the input clause used. Moreover, by the $\Delta$-Propagation Proposition B.10, $Q'_1$ is $\Delta[\tau^+]$-more general than $Q_1$.